\documentclass[prb,aps,amsmath,amssymb,twocolumn,showpacs]{revtex4}
\usepackage{graphicx}

\usepackage[toc,page]{appendix}
\usepackage{epstopdf}

\begin{document}

\title{"Creeping conductance" in nonstationary granular systems and artificial arrays}
\author{A.S.Ioselevich and V.V.Sivak}
\affiliation{L.D.Landau Institute for Theoretical Physics, Moscow
119334, Russia,\\
Moscow Institute of Physics and Technology, Moscow 141700,
Russia.}
\date{\today}

\begin{abstract}
We consider a nonstationary array of conductors, connected by  resistances that fluctuate  with time. The charge transfer between a particular pair of conductors is supposed to be dominated  by ``electrical breakdowns''  -- the moments when the corresponding resistance is close to zero.  An amount of charge, transferred during a particular breakdown, is controlled by the condition of minimum for the electrostatic energy of the system. We find the conductivity, relaxation rate, and fluctuations for such a system within the ``classical approximation'', valid,  if the typical transferred charge is large compared to $e$. We discuss possible realizations of the model for colloidal systems and arrays of polymer-linked grains.
\end{abstract}


\maketitle

\section{Introduction \label{introduction}}

Nanosize metal objects appear in many branches of modern science and technology. In microelectronics \cite{art2} they serve for creating single electron tunnelling devices \cite{GrabertDevoret} and have many optical applications \cite{optics}.     
Metal nanoparticles are also extensively used in biology and medicine \cite{art1} for tissue engineering, drug delivery, and detection of pathogens and proteins. 

A wide range of properties of such nanoparticle systems has been studied intensively, including electric conductance in both regular \cite{regular} and disordered \cite{hopping} arrays, current noise \cite{noise}, optical response, and heat transfer \cite{heat}.

One of the basic elements in most theoretical approaches to the description of the conducting grains embedded in an insulating matrix, is the Coulomb energy of the system
\begin{align}
H_C\{Q\}=\frac12\sum_{ij}U_{ij}(Q_i-q_i)(Q_j-q_j)+\sum_iQ_iV_i^{(E)}
\end{align}
where $Q_i$ are the charges of individual grains, $\hat{U}=\hat{C}^{-1}$ is the inverse of the matrix of electric induction coefficients (or capacitance matrix) $\hat{C}$, and $V_i^{(E)}$ is the external potential. In the case of homogeneous external electric field $V_i^{(E)}=-({\bf E}\cdot{\bf r}_i)$.  The ``offset charges''
$q_i$ are random variables arising due to the potentials of charged defects trapped  in the insulating matrix at random places.

Another important ingredient of the theory is the set of  tunnelling resistances $R_{ij}$ between neighbouring grains. These resistances are assumed to be large: $R_{ij}\gg R_q\equiv \hbar/e^2$; they exponentially depend on the thicknesses $d_{ij}$ of insulating layers, separating the grains:
\begin{align}
R_{ij}\sim R_0 e^{\kappa d_{ij}}
\label{coulomb2}
\end{align}
with typical value $\kappa\sim 1\;$\AA$^{-1}$. In the vast majority of papers dealing with the solid systems of nanoparticles the parameters $U_{ij}$, $q_i$, and $R_{ij}$ are assumed to be time-independent. For most systems this stationarity assumption seems to be valid --  at least as far, as robust observables, like conductivity or effective dielectric constant are discussed. For certain subtle effects, like dephasing in qubits,   which are related to very long time-scales, the fluctuations of $q_i$ due to slow migration of charged defects in the insulating matrix, are sometimes considered \cite{qubits}.

In the present paper we will be interested in the manifestly nonstationary systems, where the parameters of the network are subject to strong fluctuations in time:
\begin{align}
U_{ij}=U_{ij}(t),\quad q_{i}=q_{i}(t),\quad R_{ij}=R_{ij}(t),
\label{coulomb2a}
\end{align}
are some stochastic processes.
 Characteristic time-scale $t_0$ of these processes should not be extremely large, so that they can be relevant already for such rough effects, as the dc conductivity.

We can see three groups of systems, which, in our opinion, may satisfy the above requirements

\subsection{Colloidal solutions}

Colloidal suspensions of conducting particles exist in an abundant variety (see. e.g., \cite{colloids}).  To prevent the particles from aggregating, some sort of stabilizing agent that sticks to the particle surface is usually added to the solution, so that the particles are coated in the surfactant shells with typical thickness $\Delta d \sim 0.1 - 1$nm. Normally these shells are insulating, so that the resistance $R_{ij}$ between two particles remains relatively large even for $d_{ij}=0$, i.e., when their shells touch each other: $R_0\gg R_q$. 

Since the particles are floating in a liquid, their local environments are here, indeed, nonstationary. One can hope that the conductivity of a particular colloidal solution can be described by the model with time-dependent parameters \eqref{coulomb2a} under the following conditions:
\begin{enumerate}
\item The concentration of ions in the liquid solution should be small, so that the conduction process is not dominated by the intrinsic conductivity of the electrolyte;
\item The solution should be dense, so that the conduction process is dominated not by the motion of individual charged particles, but rather by the particle-to-particle charge transfer.
\end{enumerate}

\begin{figure}
\includegraphics[width=0.8\columnwidth]{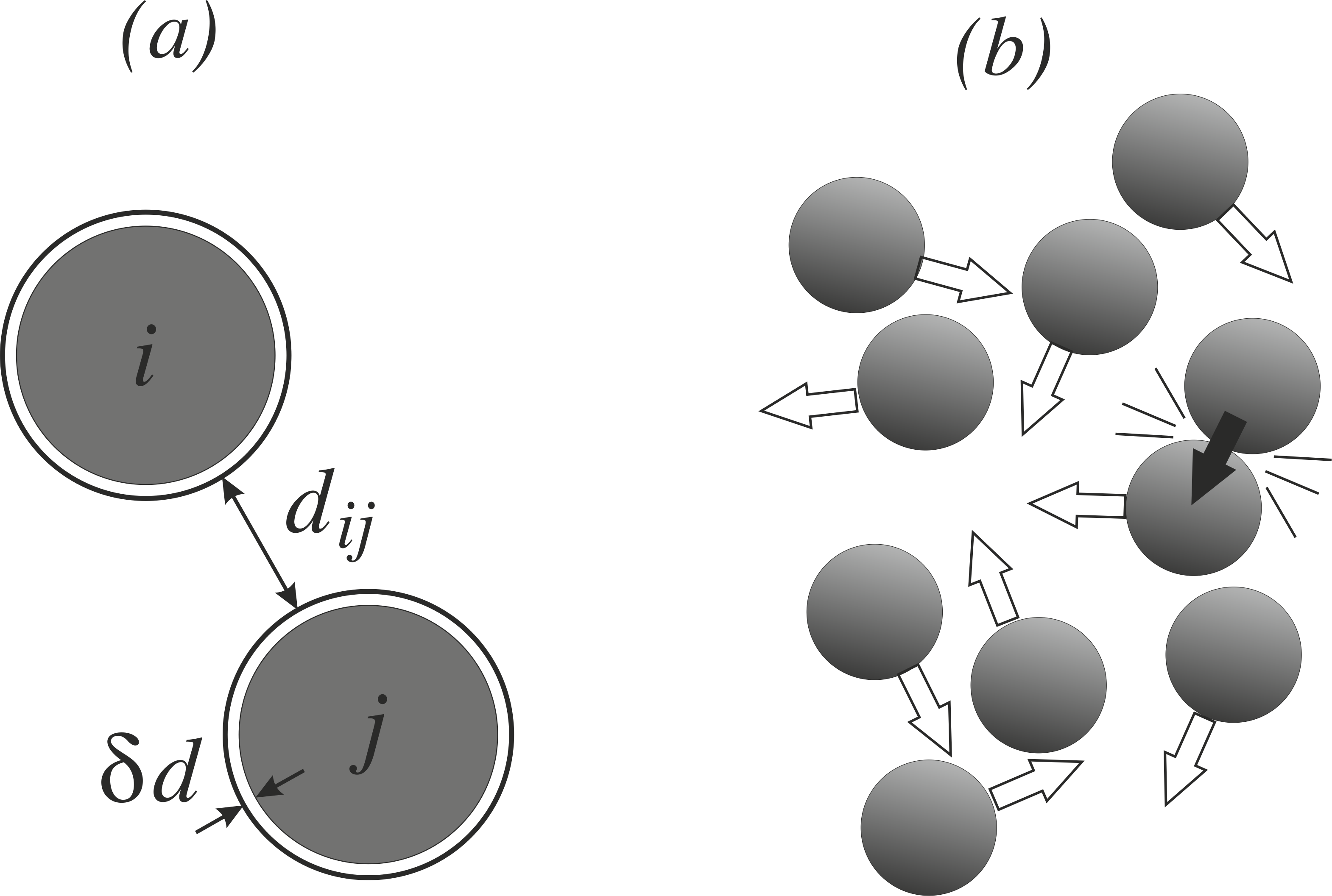}
\caption{(a) Two movable metal grains $i$ and $j$ covered with insulating coatings of thickness $\delta d$. (b) A collision of a pair of floating grains. Light arrows indicate directions of motion for individual grains, the black arrow symbolizes the pulse of current, occuring due to the collision} \label{colloid1}
\end{figure}

\subsection{Polymer-linked systems}

There is a class of artificial arrays of metal (usually -- golden) grains, connected to each other via polymer molecules (usually of the thiol family, e.g., alkanethiol CH$_3$(CH$_2$)$_n$SH, see \cite{thiol1} and references therein). The exponential dependence \eqref{coulomb2} of the intergrain resistance $R(d)$ on the length $d$ of the connecting molecule (which is proportional to $n$) is well-documented for such systems \cite{thiol1}, with $\kappa\approx 0.8\; \AA^{-1}$. The thiol molecules serve as elastic bonds, connecting massive grains, so that vibrations of the system are manifested, in particular,  as fluctuations $\delta d_{ij}$ of the bonds lengths $d_{ij}$.  Unfortunately, these bonds are relatively rigid, so that the relative amplitudes of vibrations $\varepsilon=\delta d/d$ are small. However, since the bonds are long, the parameter $\kappa d$ is large (typically $\sim 10$), and  variations of the resistances $\delta  R/R\approx (\kappa d)\varepsilon$ can be considerable.

Thus,  time dependent fluctuations of resistances may occur in polymer-linked structures. In principle, the effect can be strong, provided that soft polymer molecules are used. However, so far we were not able to find any clear experimental evidence for such an effect in the literature.

\begin{figure}
\includegraphics[width=0.8\columnwidth]{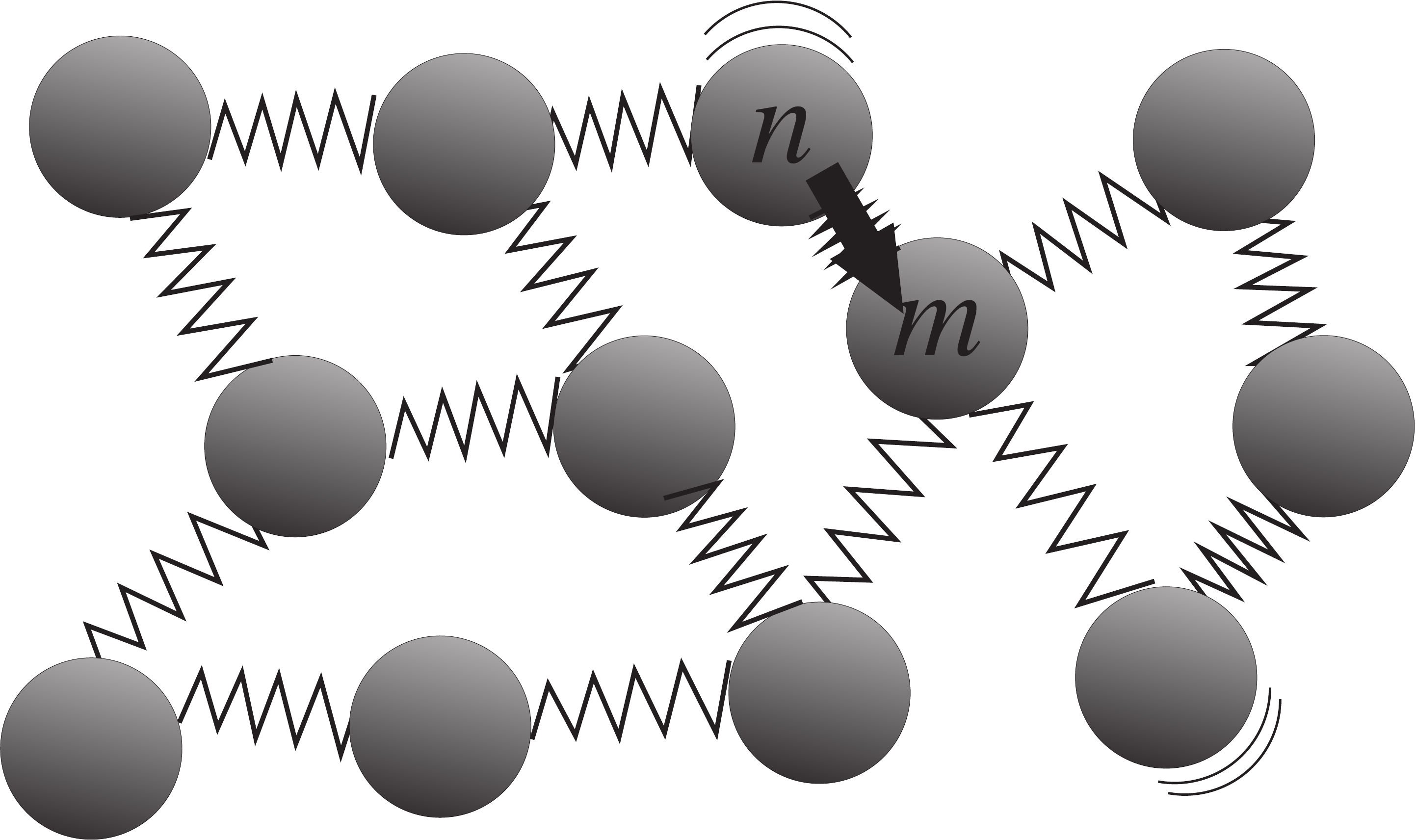}
\caption{An array of grains linked by elastic polymer molecules. Charge transfer act (shown by black arrow) occurs, when, during a vibration process, a pair of grains come especially close to each other } \label{thiol1}
\end{figure}

\subsection{Shuttled arrays}

\begin{figure}
\includegraphics[width=0.8\columnwidth]{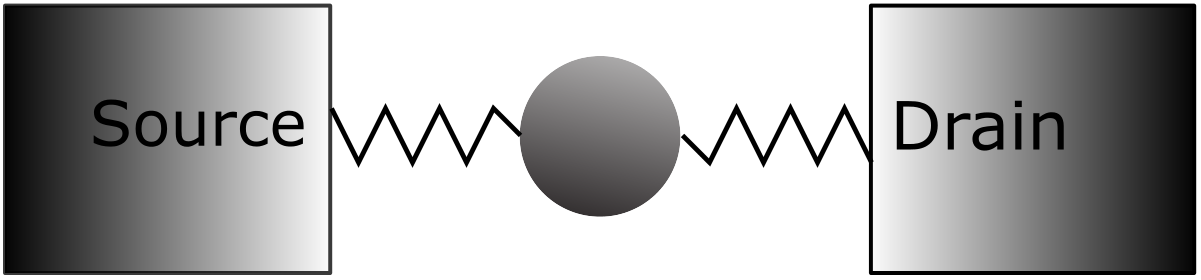}
\caption{Nanoemchanical shuttle: A grain is linked to the leads by elastic strings. } \label{shuttfig}
\end{figure}

There is a class of artificial nanodevices, namely nanomechanical shuttles \cite{shuttles-review}, that are based on a similar principle of charge transport. The simplest example of such device is nanoelectromechanical single-electron transistor (see. Fig.\ref{shuttfig}). Metallic grain is  suspended between the source and the drain by elastic strings.  Driven by Coulomb forces, the grain may approach the contacts and exchange charge with them. Thus the charge may be  transferred between the source and the drain. Due to Coulomb blockade effect it is likely that during each cycle of the grains oscillation only one electron will be exchanged, so that this shuttle serves as a single-electron tunnelling device.

Large arrays of shuttles are expected to show chaotic behaviour because of nonlinear coupling between the grains\cite{shuttles-array}. 

Since metal grains are linked by strings, they are quite movable (see. Fig.\ref{shuttfig}). The tunnelling resistances between the grains  change significantly during the vibrations, and this effect is crucial for conductive properties of a system.

\section{the model \label{the model}}

Thus, we will consider a system that can be modelled by a network with the time-dependent parameters \eqref{coulomb2a}. Moreover, we will assume that the dominant  fluctuations are those of resistance $R_{ij}$ and neglect the fluctuations of $U_{ij}$ and $q_i$. This approximation seems to be reasonable, since $R_{ij}$   exponentially depend on the fluctuating  geometrical parameters of the system (namely, on $d_{ij}$), while  for $U_{ij}$ and $q_i$  these dependences are only relatively weak power-law ones.

As to the character of fluctuations of resistances, we will adopt the following scenario:

\subsection{Requirements for the character of fluctuations}

\begin{figure}
\includegraphics[width=1\columnwidth]{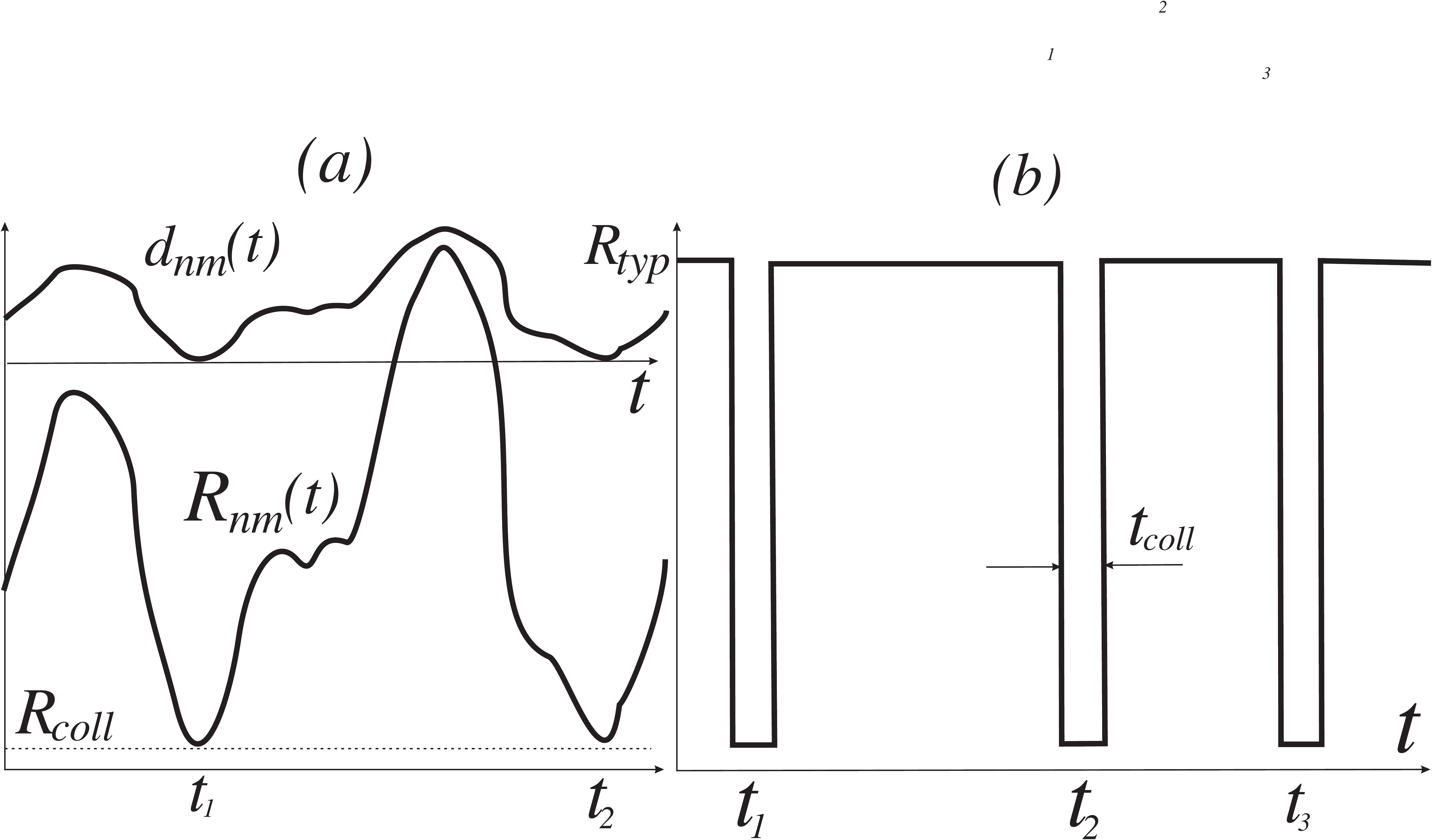}
\caption{(a) Stochastic behaviour of the distance between two grains $d_{nm}$ and  of the related resistance $R_{nm}$. The points $t_1,t_2,\ldots$, where $d_{nm}\to 0$ and $R_{nm}\to R_{\rm coll}$, correspond to ``collisions'', accompanied with breakdowns of corresponding resistances. (b) A model Poissonian process of randomly occurring  identical pulses with short duration $t_{\rm coll}$} \label{stochastic1}
\end{figure}

\begin{itemize}
\item The stochastic processes  $R_{nm}(t)$ at different bonds $\langle nm\rangle$ are statistically independent and have the same characteristics at all bonds.
\item Most of the time the resistance between each pair of neighbouring grains $\langle nm\rangle$ is quite high: $R_{nm}(t)\sim R_{\rm typ}$, but sometimes fluctuations with characteristic $R_{nm}\sim R_{\rm coll}\ll R_{\rm typ}$ occur. For simplicity we model $R_{nm}(t)$ as a Poissonian (with average frequency $w\sim 1/t_0$) sequence of pulses with duration $t_{\rm coll}\ll t_0$. The above  fluctuations are associated with the events in which the partners approach each other especially closely, so that we will call them ``collisions'' in what follows (see Fig.\ref{stochastic1}).
\item The charge relaxation time $\tau_{\rm typ}=R_{\rm typ}C$ of the pair in the ``equilibrium'' state is quite large: $\tau_{\rm typ}\gg t_0$, so that the contribution of this state to the relaxation can be neglected. Here $C$ is the characteristic scale of the capacitance ($U_{ij}\sim 1/C$). 
\item The relaxation time $\tau_{\rm coll}=R_{\rm coll}C$ of the pair in the ``collision state''  is quite short: $\tau_{\rm coll}\ll t_{\rm coll}$, there is time enough for the complete relaxation within the duration of one collision. Thus, each collision leads to effective transient electrical breakdown of corresponding resistance. During the collision, the charges are redistributed between the two grains $i$ and $j$, and the new charges $\{Q'\}$ immediately after collision may be determined from the requirement 
\begin{align}
H_C\{Q'\}\to\min
\label{coulomb2aa}
\end{align}
with additional conditions
\begin{align}
Q_n+Q_m=Q_n'+Q_m',
\quad Q_i=Q_i,\quad\mbox{for $i\neq n,m$},
\label{coulomb2ab}
\end{align}
 where $\{Q\}$ is the set of charges immediately before the collision. Naturally, all the charges after, as well as before the collision should be integers in the units of electrons charge $e$.
\end{itemize}

\subsection{Classical approximation \label{Classic approximation}}

The charge discreteness is, in general, a very important condition, leading, in particular, to the Coulomb blockade effect at low temperatures. However, we will start the study of our model from the continuous charge approximation, where  this condition is totally disregarded. This approximation can be justified, if $\overline{Q}$ -- the characteristic value of charges $Q_i$ in our problem -- is large compared to $e$. As we will see, such situation can be realised in the case of large external field $E$, applied to the system:
\begin{align}
 E\gg E_0=e(aC_{\rm gap})^{-1}.
\label{large-field1}
	\end{align}
Under this condition one can also neglect $q_i$ in the electrostatic energy, so that the condition \eqref{coulomb2aa}, governing the evolution of the charges due to a particular collision of grains $n$ and $m$, is reduced to
\begin{align}
V'_n=V'_m,\qquad V'_i\equiv V_i^{(E)}+\sum_{j}U_{ij}Q_j',
\label{coulomb2ac}
\end{align}
$V_i$ being the electrostatic potential on the $i$-th grain. The solution of \eqref{coulomb2ac} together with \eqref{coulomb2ab} gives the law of the linear transformation, which expresses the ``new'' charges $\{Q'\}$ immediately after the collision through the ``old'' ones $\{Q\}$ that existed immediately before the collision: 
\begin{align}
\label{capp5}
Q'_i=Q'_i\{Q\}=\sum_{j}g_{ij}^{\langle nm\rangle}Q_j+X_i^{\langle nm\rangle},
\end{align}
with
\begin{align}
g_{ij}^{\langle nm\rangle}=\delta_{ij}-C_{\rm eff}^{\langle nm\rangle}(\delta_{in}-\delta_{im})(U_{nj}-U_{mj}),
\label{capp6}
\end{align}   
\begin{align}
\label{capp6qw}
X_i^{\langle nm \rangle}=-C_{\rm eff}^{\langle nm\rangle}(\delta_{in}-\delta_{im})(V^{(E)}_n-V^{(E)}_m),
\end{align}
\begin{align} 
C_{\rm eff}^{\langle nm\rangle}=(U_{nn}+U_{mm}-2U_{mn})^{-1}.
\end{align}     
The upper indices $\langle nm\rangle$ in \eqref{capp5} indicate that the collision occurs between grains $n$ and $m$. Note, that in an important case of a regular array with translational symmetry, considered in the following sections,  $C_{\rm eff}^{\langle nm\rangle}\equiv C_{\rm eff}$ does not depend on the position of the bond.

\begin{figure}
\includegraphics[width=0.8\columnwidth]{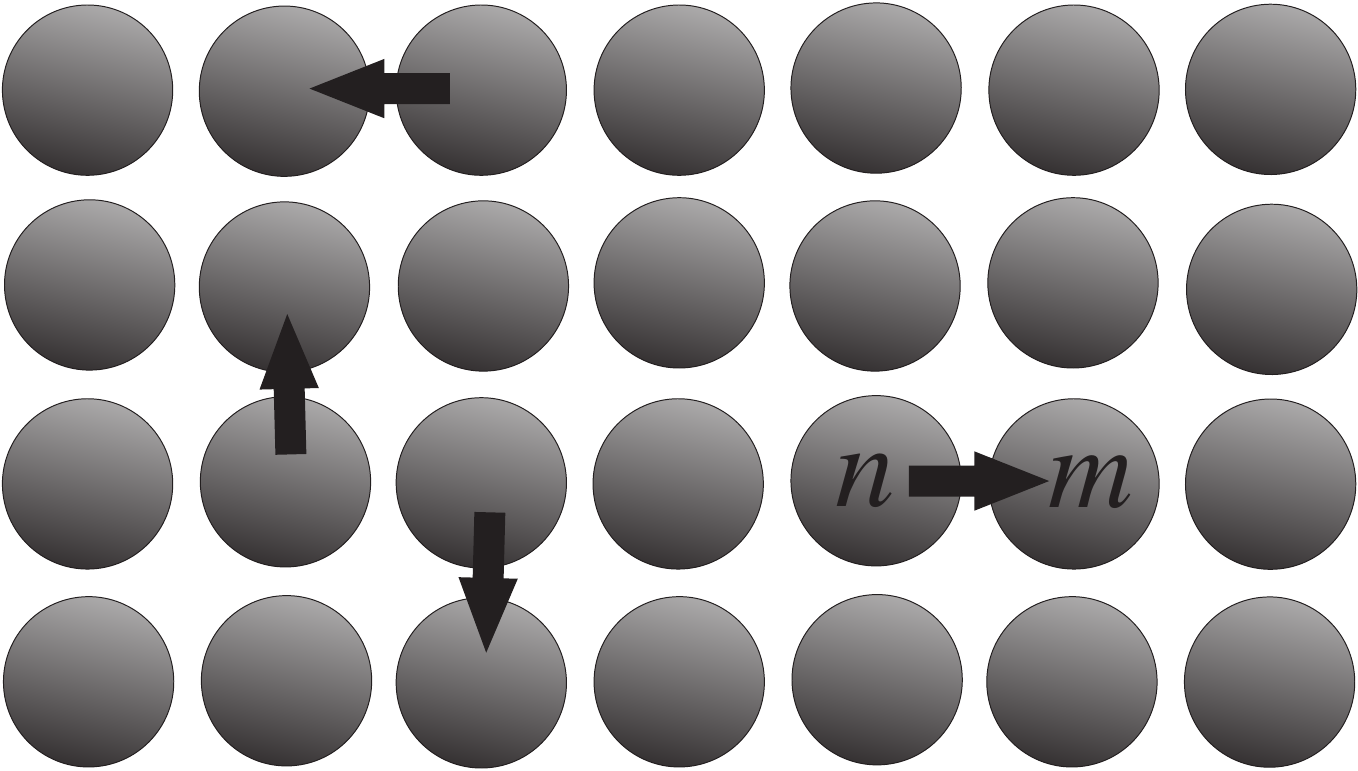}
\caption{The stochastic process of charge transfer in a regular array of grains. Black arrows show pulses of current, occurring at random bonds in random moments of time} \label{array1}
\end{figure}

Thus, the evolution of the system in the classical approximation is reduced to subsequent application of the linear transformations \eqref{capp5}, occurring at random moments at randomly chosen bonds  $\langle nm\rangle$ (see Fig.\ref{array1}). These transformations describe partial equilibriums of the system taking place during the collisions.

\section{Electrostatics of a regular array}

Although any realistic network of grains should be to some extent disordered,   we expect that in many cases this geometric disorder does not  lead to any dramatical changes
of the effects, caused by the collision-like fluctuations, described in the previous subsection. So
 in this paper we restrict our consideration to an infinite {\it regular array of identical conducting grains}. The grains are assumed to be placed close to each other, so that the size of the grains is approximately equal to the distance $a$ between neighbouring grains, while the minimal width $d$ of the insulating gap between neighbouring grains is narrow: $d\ll a$. Under these condition  the matrix $\hat{C}$ is dominated by capacitances of these gaps $C_{\rm gap}$ which are larger than the geometric capacitance $C_{\rm sol}\propto a$ of a solitary grain
\begin{align}
 \Lambda=C_{\rm gap}/C_{\rm sol}\gg 1.
 \label{cap2}
	\end{align}	
For example, if the grains are identical spheres, then\cite{two spheres} 
\begin{align}
C_{\rm gap}^{\rm (sph)}\approx(\varepsilon_{\rm gap}a/8)\ln\left(a/d\right),\quad C_{\rm sol}^{\rm (sph)}\approx \varepsilon_{\rm out}a/2.
\label{cap3}
	\end{align}	
Here $\varepsilon_{\rm gap}$ is the dielectric constant of the insulating gap between the grains and $\varepsilon_{\rm out}$ is the dielectric constant of the ``outer space'' -- the three-dimensional medium in which the array is embedded. In the case of a three dimensional array one should simply put $\varepsilon_{\rm out}=\varepsilon_{\rm gap}$, but for low dimensional arrays these two constants may be different (see below). 

In realistic cases, the actual value of $C_{\rm gap}$  strongly depends on the geometry of the system. Since we are not going to restrict our consideration by some definite choice of the grains shape, we treat $C_{\rm gap}$ as an independent phenomenological characteristic of our system in what follows.

\subsection{The leading approximation}

In the leading approximation in parameter $\Lambda\gg 1$ one can neglect all geometric capacitances. In this approximation the total charge $Q_i$ of any grain $i$ is split into $z$ ($z$ is the coordination number of our array) parts $q_{ij}$, each of them being localised near the contact with neighbour $j$. Then
\begin{align}
q_{ij}=C_{\rm gap}(V_i-V_j),\quad
Q_i=\sum_{j:\langle ij\rangle}q_{ij},
\label{cap4}
	\end{align}	
where the summation runs over all $z$ neighbours $j$ of the grain $i$. So, one readily gets the capacitance matrix
\begin{align}
C_{ij}=-C_{\rm gap}\Delta_{ij},\quad \Delta_{ij}=\sum_{n:\langle in\rangle}(\delta_{jn}-\delta_{ji})
\label{cap5}
	\end{align}	
The matrix $\Delta_{ij}$ is often called ``discrete Laplace operator'' in discrete mathematics. The translational invariance of the regular lattice leads to the relation $C_{ij}=C_n$, where $n\equiv i-j$ and ${\bf r}_n$ is a radius-vector, connecting sites $i$ and $j$. In the Fourier domain
\begin{align}
C({\bf k})=\sum_n C_{n}e^{i({\bf k}\cdot{\bf r}_n)}=C_{\rm gap}\sum_{\vec{\delta}}(1-e^{i({\bf k}\cdot\vec{\delta})})=\nonumber\\=4C_{\rm gap}\sum_{\mu}\sin^2(ak_\mu)=\frac{8C_{\rm eff}}{z}\sum_{\mu}\sin^2(ak_\mu),
\label{cap5a}
	\end{align}	
where the summation in the first line runs over all $z$ vectors $\vec{\delta}$, connecting certain site of the lattice with its nearest neighbours, while in the second line -- over $z/2$ spatial directions $\mu$. Thus, the function $U_{n}$ should obey the  equation $-C_{\rm gap}\hat{\Delta}\cdot\hat{U}=1$ from which the Gauss theorem immediately follows. The latter theorem, in particular, establishes a simple relation between the constant $C_{\rm eff}$, entering \eqref{capp6} and  \eqref{capp6qw}, and the gap capacitance 
$C_{\rm gap}$:
\begin{align}
C_{\rm eff}^{\langle nm\rangle}=C_{\rm eff}=\frac{1}{2(U_0-U_1)}\equiv zC_{\rm gap}/2,
\label{cap0rel1}
	\end{align}	
where $z$ is the coordination number of the lattice.

For large-distance tails of $\hat{U}({\bf r})$ at $r\gg a$ we get:  
\begin{align}
 U({\bf r})=\int\frac{d^D{\bf k}}{(2\pi)^D}\frac{e^{-i({\bf k}\cdot{\bf r})}}{C({\bf k})}\approx\nonumber\\\approx \frac{1}{\tilde{C}_{\rm gap}}\left\{\begin{aligned}b^{(1D)}-r/2a,&\quad\mbox{for $D=1$},\\
 b^{(2D)}-\frac{\ln(r/a)}{2\pi},&\quad\mbox{for $D=2$},\\
\frac{a}{4\pi r}, &\quad\mbox{for $D=3$}.
 \end{aligned}\right.
\label{green1pw}
	\end{align}
Unknown constants $b^{(1D)},b^{(2D)}$	appearing in the low-dimensional versions of \eqref{green1pw} arise due to pathological divergence of $U({\bf r})$ at large distances in low dimensions (see below). 
In three-dimensional case the divergence does not occur, and we get a very natural result
\begin{align}
 U^{(3D)}({\bf r})\approx\frac{1}{\varepsilon_{\rm eff}r},\label{green1ppw}\\
  \varepsilon_{\rm eff}=\frac{4\pi C_{\rm gap}}{a}=4\pi\Lambda\varepsilon_{\rm out}.
\label{green1ppw0}
	\end{align}
which describes effective dielectric screening of the Coulomb potential.

\subsection{Inconsistency of the leading approximation for low-dimensional arrays}

The three-dimensional version \eqref{green1ppw} of the result \eqref{green1pw} remains valid for arbitrary large distances $r$, provided there is no Debye screening in the system (see below). It is not the case, however,  as long as low-dimensional arrays are concerned. Indeed, the 1D and 2D versions of the result \eqref{green1pw} can not be valid at large enough distances  because of the imperative  requirement $U>0$  (see, e.g \cite{theorem}). It also shows that zero approximation in small parameter $\Lambda^{-1}$, which we have used above, becomes insufficient at large distances in low dimensions, and small corrections to $\hat{C}$ should be taken into account, which somehow will be crucial at large distances and will resolve the problem:
\begin{align}
C_{ij}=-C_{\rm gap}\Delta_{ij}+\delta C_{ij},\quad \delta C_{ij}\sim C_{\rm sol}\ll C_{\rm gap}.
\label{cap5h}
	\end{align}	
The result  \eqref{green1pw} actually implies the true $D$-dimensional electrostatics, in which the field lines are supposed to be confined to the array. For $D<3$ it is  important to have in mind that our low-dimensional array is embedded in the real three-dimensional space (with a dielectric constant $\varepsilon_{\rm out}$) and the field lines would eventually escape the array at large enough distances. As a consequence the dielectric screening of Coulomb potential due to polarisation of the grains becomes irrelevant at these large distances and the conventional bare Coulomb potential is restored, so that the law \eqref{green1pw} is substituted by 
\begin{align}
 U({\bf r})\approx\frac{1}{\varepsilon_{\rm out}r}.
\label{green1pp}
	\end{align}
	
	\subsection{Metal gate: Debye screening}
	
The above problem of inconsistency is often  resolved in a somewhat voluntary way, just by choosing some simplistic form of correction, usually the self-capacitance one: $\delta C_{ij}^{(0)}=C_0\delta_{ij}$. In the ${\bf k}$-representation it reads
\begin{align}
C({\bf k})\approx C_{\rm gap} (ka)^2 +C_0.
\label{cap7}
	\end{align}	
However, any $C({\bf k})$ with $C(0)\neq 0$ leads to an exponential decay 
\begin{align}
U({\bf r})\propto e^{-r/r_D},\qquad r\gg r_D=a(C_{\rm gap}/C_0)^{1/2}
\label{cap7s}
	\end{align}	
$r_D$ being the effective screening radius.  Such a decay can only be physically justified for systems with Debye screening,  caused  by  presence of a metal gate.
Note that the gate leads to the Debye screening also for a three-dimensional array. 

\subsection{No metal gate: dielectric screening in low dimensions}

If we assume that our system is a global insulator, then the requirement 
\begin{align}
\lim_{k\to 0}\delta C({\bf k})=0,
\label{cap8}
	\end{align}	
must be fulfilled. This requirement, in its turn, indicates that the $\delta C({\bf k})$ should be non-analytic function of $k$ at $k=0$ in low dimensional arrays. Indeed the only analytic behaviours, compatible with the rotational symmetry would be either $\delta C({\bf k})\approx {\rm const}$ (the metallic case, considered in the previous subsection), or  $\delta C({\bf k})\propto k^2$ (an irrelevant small renormalisation of the constant $C_{\rm gap}$).

Now we will present physical arguments that allow for reconstruction of  $\delta C^{(0)}({\bf k})$ in the insulating case on a quantitative level.

Since the true long-distance asymptote of $U({\bf r})$ is given by \eqref{green1pp}, we can find the low-${\bf k}$ behaviour of $C({\bf k})$, or, rather of  $\delta C({\bf k})$, by means of Fourier-transforming \eqref{green1pp}:
 \begin{align}
 \delta C({\bf k})=\left(\int \frac{a^Dd^D{\bf r}}{\varepsilon_{\rm out}r}e^{i({\bf k}\cdot{\bf r})}\right)^{-1}\approx\nonumber\\\approx \varepsilon_{\rm out}a\left\{\begin{aligned}-\frac{1}{2\ln(|k|a)},&\quad\mbox{for $D=1$},\\
 |k|a/2\pi,&\quad\mbox{for $D=2$},
 \end{aligned}\right.
\label{correction-k}
	\end{align}
	and, finally
 \begin{align}
 C({\bf k})=C_{\rm gap}\left\{\begin{aligned}(ka)^2-\frac{1}{\Lambda\ln(|k|a)},&\quad\mbox{for $D=1$},\\
 (ka)^2+\frac{|k|a}{\pi\Lambda},&\quad\mbox{for $D=2$}.
 \end{aligned}\right.
\label{correction-k1}
	\end{align}
Thus, we have shown that the corrections are indeed non-analytic and indeed do dominate at small $k$ in both two- and one-dimensional arrays. 

It should be stressed that the basic result \eqref{correction-k1} in the form of a sum of two contributions is not just an interpolation formula between two limiting cases: it is quantitatively valid also in the case when both terms are of the same order of magnitude, the only condition for its validity being $k\ll a^{-1}$. Indeed, the formula  \eqref{correction-k1} represents the first two terms in the expansion of $C({\bf k})$ in series of powers of small parameter $\Lambda^{-1}$. These coefficients are functions of the parameter $ka$ only, and we have found their asymptotic expressions for $ka\ll 1$.
Although the coefficient in the second term was found with the help of physical arguments, applicable only in the case when this second term is dominant, the obtained result for this coefficient must be valid for $ka\ll 1$, independent on which term of the two dominates at given value of $\Lambda$.

In order to highlight the dimensional crossover, occurring with increase  of distance, it is instructive to find $C({\bf r})$ and $U({\bf r})$ using the expressions \eqref{correction-k1}. We will do that for the 2D and 1D cases separately.

\subsection{Two-dimensional array:  crossover from 2D to 3D electrostatics \label{Two-dimensional array}}

Making the inverse Fourier transformation of the lower line of \eqref{correction-k} we come to 
\begin{align}
 C^{(2D)}({\bf r})=\frac{C_{\rm sol}a^3}{\pi}\int \frac{d^2{\bf k}}{(2\pi)^2}|{\bf k}|e^{-i({\bf k}\cdot{\bf r})}=-\frac{C_{\rm sol}}{2\pi^2}\left(\frac{a}{r}\right)^3,
\label{correction-kyt}
	\end{align}
	valid for $r\gg a$. Thus, the tail of $C^{(2D)}({\bf r})$ at large $r\gg a$ is completely dominated by the correction term, proportional to $C_{\rm sol}$. Note also, that
the off-diagonal elements of the matrix $C_{ij}$ are negative, in accord with the general requirement (see \cite{theorem}). Similarly, from the lower line of \eqref{correction-k1} we obtain
\begin{align}
 U^{(2D)}({\bf r})=\frac{1}{C_{\rm gap}}F^{(2D)}\left(r/r_c^{(2D)}\right),
 \label{correction-kyt1}
	\end{align}
where 
\begin{align}
 r_c^{(2D)}=\pi a\Lambda\gg a,
 \label{length1}
	\end{align}
is the characteristic length-scale of crossover, and the function
\begin{align}
F^{(2D)}(z)\equiv\frac{1}{2\pi}\int_0^{\infty}J_0(zt)\frac{dt}{t+1}=\frac{1}{4}[{\bf H}_0(z)-Y_0(z)]
 \label{length2}
	\end{align}
can be expressed in terms of the Struve function ${\bf H}_0(z)$ and the Neumann function $Y_0(z)$. This result was obtained earlier by J.~E.~Mooij and G.~Schoen (see \cite{Mooij}), who used the analogy with the problem of  screening of the Coulomb field of a point-like charge, placed in an insulating film, solved by L.~V.~Keldysh \cite{Keldysh} 

\begin{figure}[h]
\includegraphics[width=0.9\columnwidth]{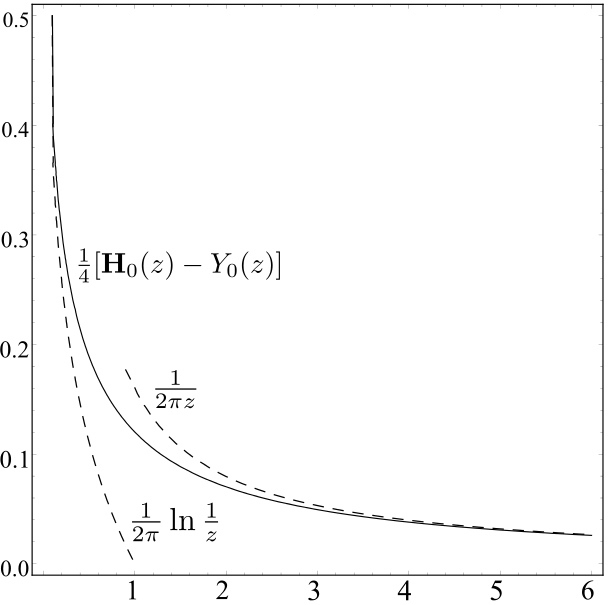}
\caption{Plot of the function $F(z)$, defined by \eqref{length2}. Its asymptotical forms for small and large values of the argument  are shown by dashed lines.} \label{struve}
\end{figure}

The general shape of the function $F$ is shown in Fig.\ref{struve} Its asymptotes 
\begin{align}
F^{(2D)}(z)\approx\frac{1}{2\pi}\left\{\begin{aligned}\ln(1/z),&\quad\mbox{for $z\ll 1$},\\
 1/z,&\quad\mbox{for $z\gg 1$}.
 \end{aligned}\right.
 \label{length3}
	\end{align}
reproduces the expression \eqref{green1pp} (lower line) and  the 2D-version of expression \eqref{green1pw} (upper line).  Moreover, comparing \eqref{length3} with  \eqref{green1pw}, one can identify the unknown constant $b^{(2D)}$ entering Eq.\eqref{green1pw}:
\begin{align}
b^{(2D)}=\frac{1}{2\pi}\ln\Lambda\gg 1.
 \label{length2d}
	\end{align}
 Thus, the result \eqref{correction-kyt1} describes smooth crossover from the 2D electrostatics, confined to the array, to the conventional 3D one, which sets on, when spatial scale of the problem exceeds the crossover radius $r_c^{(2D)}$. The repulsion energy of two electrons, placed at distance $r$ from each other
 \begin{align}
V(r)=E_CF^{(2D)}(r/r_c^{(2D)}), \quad E_C=e^2/2C_{\rm gap}.
	\end{align}
 is equal to $E_C\ln\Lambda/2\pi$ for nearest neighbours ($r=a$) and decreases only very slowly up to $r\sim r_c^{(2D)}$, where the conventional 3D   Coulomb law $V(r)=e^2/2\varepsilon_{\rm out}r$ sets on.

\subsection{One-dimensional array:  crossover from 1D to 3D electrostatics \label{One-dimensional array}}

Following the same lines, as in the previous subsection, we get the long-distance tail of $C_{ij}$:
\begin{align}
 C^{(1D)}({\bf r})=-C_{\rm sol}a\int_{-\infty}^{\infty} \frac{d k}{2\pi}\frac{e^{-ikr}}{\ln(|k|a)}\approx-\frac{C_{\rm sol}}{2(r/a)\ln^2(r/a)},
\label{correction-kyl}
	\end{align}
\begin{align}
 U^{(1D)}({\bf r})\approx\frac{1}{\varepsilon_{\rm out}r_c^{(1D)}}F^{(1D)}(r/r_c^{(1D)},\ln\Lambda).
\label{green1q171m}
	\end{align}
	where		the function of two variables 
	\begin{align}
F(z,y)\approx\frac{1}{\sqrt{z^2+1}}+\frac{y}{2}e^{-z}
	\end{align}
and the crossover radius is
\begin{align}
 r_c^{(1D)}=a\left(\Lambda\ln\Lambda/2\right)^{1/2}\gg a.
 \label{length1f}
	\end{align}
Derivation of this formula is given in Appendix \ref{Evaluation of in one-dimensional case}. The relevant asymptotic forms of this function are:
\begin{align}
F(z,y)\approx\left\{\begin{aligned}\frac{y}{2}(1-z),&\quad\mbox{for $z\ll 1$, $y\gg 1$},\\
 1/z,&\quad\mbox{for $z\gg \ln y$, $y\gg 1$}.
 \end{aligned}\right.
\label{asy9x}
	\end{align}
The upper line of \eqref{asy9x} reproduces the result \eqref{green1pw} with the constant $b^{(1D)}\approx\left(\Lambda\ln\Lambda/2\right)^{1/2}/2$. The lower line reproduces the bare three-dimensional Coulomb interaction \eqref{green1pp}.

\section{the charge and current relaxation}

In this Section we discuss the evolution of the distribution of the charges in the system in the absence of external field (for $V^E_i=0$). Under this condition the linear transformation \eqref{capp5} is homogeneous, and one can write 
\begin{align}
\label{gf1}
Q_i(t)=\sum_jG_{ij}(t)Q_j(0), \quad \hat{G}(t)=\prod_{0<t_\lambda<t}\hat{g}^{\langle nm\rangle_\lambda}
\end{align} 
where the index $\lambda$ numerates the collisions. 
It is natural to call the matrix $\hat{G}(t)$ the evolution operator, or the Green's function.

Clearly, $\hat{G}$ depends on the specific sequence  of collisions and, therefore, is a random operator. In order to determine the statistically averaged charge distribution we have to find $\overline{\hat{G}}(t)$. Since the collisions happen at random places without any correlation, the individual factors $\hat{g}^{\langle nm\rangle_\lambda}$ in the product in \eqref{gf1} can be averaged over the position of bond $\langle nm\rangle$ independently. As a result, we obtain
\begin{align}
\overline{\hat{G}}=\langle\hat{g}^M\rangle_M, 
\end{align}
where
\begin{align}
\label{gf4}
\hat{g}=\langle\hat{g}^{\langle nm\rangle}\rangle_{\langle nm\rangle}=1+\frac{C_{\rm eff}}{N_b}\hat{\Delta}\cdot\hat{U}
\end{align} 
is the matrix   $\hat{g}^{\langle nm\rangle}$, averaged over all possible positions of the bond $\langle nm\rangle$, 
and $N_b$ is a total number of  bonds (pairs of neighbouring grains) in the system.
The random variable $M$ is a total number of collisions that have occurred within the time interval $(0,t)$. Its average is $\overline{M}=N_bwt$ and its dispersion is relatively small:
\begin{align}
\label{gf2}
 \left(\overline{\left(M-\overline{M}\right)^2}\right)^{1/2}\sim \overline{M}^{1/2}\ll \overline{M},
\end{align}
so that at $N_b\to\infty$ the fluctuations of $M$ can be neglected and
\begin{align}
\label{gf3}
\overline{\hat{G}}=\hat{g}^{\overline{M}}=\left(1+\frac{C_{\rm eff}}{N_b}\hat{\Delta}\cdot\hat{U}\right)^{N_bwt}\approx\exp\left\{C_{\rm eff}wt\hat{\Delta}\cdot\hat{U}\right\}
\end{align}

Formulas \eqref{gf4}, \eqref{gf2} and \eqref{gf3} lead us to the following general expression for the average evolution operator, valid for arbitrary $U_{ij}$:
\begin{align}
\overline{G}_{ij}(t)=\int_{\rm BZ}\frac{a^Dd^D{\bf k}}{(2\pi)^D}\exp\left\{C_{\rm eff}\Delta({\bf k})U({\bf k})wt-i{\bf k}\cdot{\bf r}_{ij}\right\}
\label{gf6}
\end{align}
where integration should be held over the Brillouin zone. 

For the current through  bond $\langle nm\rangle$ (in the direction from $n$ to $m$) one can write
\begin{align}
I_{nm}(t)=\sum_{t_\alpha}\tilde{\delta}(t-t_{\alpha}) \{Q_m(t_\alpha+0)-Q_m(t_\alpha-0)\}=\nonumber\\=C_{\rm eff}\sum_{t_\alpha}\tilde{\delta}(t-t_{\alpha})\sum_k(U_{n-k}-U_{m-k})Q_k(t_\alpha-0),
\label{co6as}
	\end{align}
	where the summation runs over moments $t_\alpha$ corresponding to the $\langle nm\rangle$-collisions. The current consists of a sequence of short pulses with a shape $\tilde{\delta}(t-t_{\alpha})$ which reproduces the shape of the pulses in $R_{nm}(t)$. The intensities of the pulses are linear functions of the charges just before the collision. By means of averaging of \eqref{co6as} it can be shown that the average currents at the moment $t$ can be expressed through the average charges at the same moment:
	\begin{align}
\overline{I}_{\langle nm\rangle}(t)=wC_{\rm eff}\sum_{i}(U_{ni}-U_{mi})\overline{Q}_{i}(t)=\nonumber\\=R_{\rm eff}^{-1}(\overline{V}_m(t)-\overline{V}_n(t)),
\label{aver-cur}
	\end{align}
where 
	\begin{align}
R_{\rm eff}=(wC_{\rm eff})^{-1},
\label{gr7uq}
	\end{align}
is the effective time-averaged resistance of a bond. 

So, on average, our system behaves as a regular lattice, where each pair of neighbours is connected by  an ohmic resistance $R_{\rm eff}$. If we place a charge $Q_i$ in some site $j$ of this regular lattice, it produces long-range fields
	\begin{align}
E_{\langle nm\rangle}=(U_{n-i}-U_{m-i})(Q_i/a),
\label{fie1}
	\end{align}
which, in their turn, give rise to currents in the whole sample (not only in the immediate vicinity of the charge).

\subsection{The leading approximation: homogeneous relaxation}

In the leading approximation in $\Lambda^{-1}$, when the expression \eqref{cap5} is valid, the result \eqref{gf3} is dramatically simplified:
\begin{align}
\label{mr1}
\overline{G}_{ij}(t)=e^{-t/\tau}\cdot\delta_{ij}, \qquad\tau=2/zw.
\end{align}
So, the average charge at the initial site decays exponentially, while  at all other sites average charges remain zeros. For an arbitrary initial distribution of charges $Q_i(0)$ the homogeneous relaxation is predicted:
\begin{align}
\label{mr1a}
\overline{Q}_{i}(t)=e^{-t/\tau}Q_i(0).
\end{align}

 At the first glance this result seems to contradict the total charge conservation, but the contradiction is resolved, if one looks at the divergence of the average current density at site $n$, induced by a charge $Q_i$, placed at site $i$:
 \begin{align}
\label{diver1}
\frac{\partial Q_n(t)}{\partial t}=-R_{\rm eff}\left.(\nabla\cdot {\bf E}(t))\right|_n=\nonumber\\=R_{\rm eff}\left[\hat{\Delta}\cdot\hat{U}\right]_{ni}Q_i(t)=-\delta_{ni}\frac{Q_i(t)}{\tau}.
\end{align}
It means that the average charges at all sites, except the origin, do not change with time and remain zeros.
There is a full analogy to a classical problem of relaxation for a charged sphere, placed into an infinite conducting medium. In this problem the initial charge also decays with time exponentially, due to currents, arising in the medium. These currents, produced by the long-range Coulomb fields, deliver the decaying charge {\it directly to infinity}, while no charge can be detected at any  intermediate place at finite distance from the sphere. Certainly, this result is valid only in a quasistatic approximation, when the velocity of light is assumed to be infinite and no retardation effects are taken into account.

Thus, we conclude that in the leading approximation in $\Lambda^{-1}$ the average charge goes to spatial infinity without being trapped anywhere.

In three-dimensional arrays this result holds also beyond the leading approximation, provided there is no Debye screening. 

\subsection{Metal gate: relaxation via diffusion}

In the presence of the Debye screening the relaxation becomes inhomogeneous in arrays of all dimensions. In this subsection we will consider the case of a three-dimensional array, where the screening is the only mechanism that gives rise to the inhomogeneity. On the other hand, in low dimensional arrays there is a competing mechanism due to modification of dielectric screening at large distances $r\gtrsim r_c$. The latter mechanism dominates, if $r_c<r_D$. 

In 3D case we have
\begin{align}
	U_{ij}=e^{-r/r_D}/\varepsilon_{\rm eff}r,\qquad r\equiv r_{ij},
\label{no-scr1a}
	\end{align}
and the general expression \eqref{gf6} can be, for large distances $r_{ij}\gg a$, rewritten as
\begin{align}
\overline{G}_{ij}(t)\approx\int\frac{a^3d^3{\bf k}}{(2\pi)^3}\exp\left\{-\frac{t}{\tau}\frac{(kr_D)^2}{(kr_D)^2+1}-i{(\bf k}\cdot{\bf r}_{ij})\right\}
\label{gr7aea}
	\end{align}
	This integral can be evaluated in different limiting cases.

At initial stage of relaxation 	($t\lesssim \tau$) the average charge $\overline{G}_{ij}(t)$ at site $i\neq j$ is arising mainly due to the field produced by the initial unit charge at point $j$:
\begin{align}
\overline{G}_{ij}(t)\approx tR_{\rm eff}\left[\hat{\Delta}\cdot\hat{U}\right]_{ij}=\frac{t}{\tau}\frac{a^3}{4\pi r_D^2 r} e^{-r/r_D}
\label{gr7aea1}
	\end{align}
	During this early stage a finite part ($\sim t/\tau$) of the initial charge leaves the origin and spreads over the domain of size $\sim r_D$, while only exponentially small fraction of the charge reaches distances $r\gg r_D$.
	
	At late stage of relaxation $t\gg \tau$ almost all the charge is distributed over the domain of size
\begin{align}
\overline{r}(t)=(D_{\rm eff}t)^{1/2}\sim r_D(t/\tau)^{1/2}\gg r_D.
\label{gr7aea2}
	\end{align}
	On this stage the secondary charge is produced not by the field of the initial charge at the origin, but by the secondary charges themselves, in a self-consistent way. It results in an effective diffusion of charge:
\begin{align}
\overline{G}_{ij}(t)\approx \frac{a^3}{(4\pi D_{\rm eff}t)^{3/2}}\exp\left\{-\frac{r^2}{4D_{\rm eff}t}\right\}, \quad D_{\rm eff}=\frac{r_D^2}{\tau}.
\label{gr7aea3}
	\end{align}
It should be noted that the diffusive law \eqref{gr7aea3} ceases to work at very large distances $r\gg \overline{r}(t)$, where the distribution \eqref{gr7aea1} remains valid even at $t\gg \tau$. This far tail of the distribution, however, describes only an exponentially small fraction of the total charge.

\begin{figure}[h]
\includegraphics[width=0.8\columnwidth]{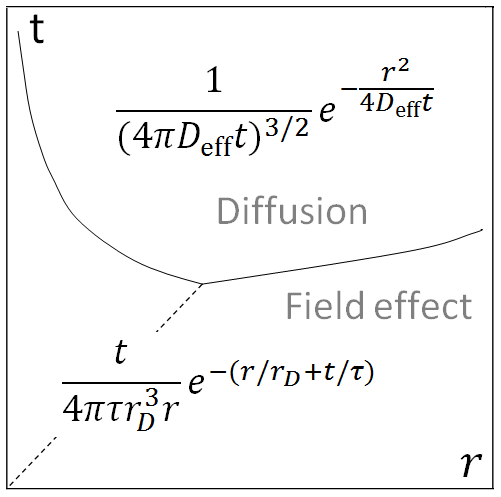}
\caption{Different modes of relaxation in a 3D array with weak Debye screening (arising, e.g., due to presence of a metal gate). For small $t$ and/or $r$ the charge relaxes due to the field effect -- here the charges are small, though the currents are large. For large $t$ and/or $r$ the standard diffusion sets on.} \label{3d}
\end{figure}

The phase diagram, depicting different modes of relaxation on the $r-t$ plane is shown in Fig.\ref{3d} 
	
\subsection{Inhomogeneous relaxation in two-dimensional arrays}

 As we have mentioned, for low-dimensional arrays the relaxation becomes inhomogeneous at large times and distances already without Debye screening, just because of the existence of the crossover radius $r_c$. 
 
\begin{figure}[h]
\includegraphics[width=0.8\columnwidth]{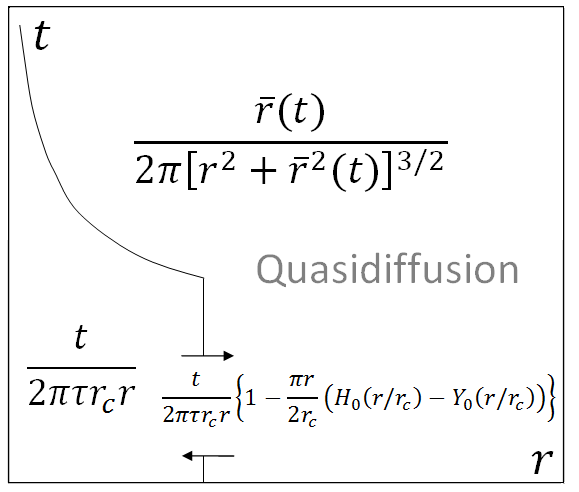}
\caption{Different modes of relaxation in a 2D array with large $r_c\gg a$. As in the weakly screened 3D case, for small $t$ and/or $r$ the charge relaxes due to the long range field effect. For large $t$ and/or $r$ the  quasi-diffusion regime, described by the law \eqref{exp2f}, sets on.} \label{2d}
\end{figure}

 At large distances $r\gg a$ we can write
\begin{align}
\overline{G}_{ij}(t)\approx\int\frac{a^2d^2{\bf k}}{(2\pi)^2}\exp\left\{-\frac{t}{\tau}\frac{kr_c}{1+kr_c}-i({\bf k}\cdot{\bf r}_{ij})\right\}
\label{2dd0}
	\end{align}
	Again, as in the previous subsection, we consider separately the early ($t\lesssim\tau$) and the late ($t\gg\tau$) stages of the relaxation.
	At the early stage
\begin{align}
\overline{G}_{ij}(t)\approx tR_{\rm eff}\left[\hat{\Delta}\cdot\hat{U}\right]_{ij}=\nonumber\\=\frac{t}{\tau}\frac{a^2}{2\pi rr_c}\left\{1-\pi (r/2r_c)[{\bf H}_0(r/r_c)-Y_0(r/r_c)]\right\}\approx\nonumber\\\approx \frac{t}{\tau}\left\{\begin{aligned}\frac{a^2}{2\pi rr_c},\qquad & (a\ll r\ll r_c),\\
\frac{r_ca^2}{2\pi r^3},\qquad & (r\gg r_c=r_c^{(2D)}\equiv\pi a\Lambda).
\end{aligned}\right.
\label{gr7aea1we}
	\end{align}
	At the late stage, as in the case with the Debye screening, the secondary charges are produced in a self-consistent way, but here it leads not to the conventional diffusion of the charge, but to some kind of ``quasidiffusion'', described by the evolution law
\begin{align}
\overline{G}_{ij}(t)\approx \frac{\overline{r}(t)a^2}{2\pi\left[r^2+\overline{r}^2(t)\right]^{3/2}},\qquad  \overline{r}(t)=r_ct/\tau.
\label{exp2f}
	\end{align}
Thus, at large enough times ($t\gg\tau$) the whole (with the accuracy up to exponentially small in $t/\tau$ corrections) charge is distributed over the area with linear size $\overline{r}(t)$. In contrast with the case of standard diffusion, this size grows linearly with $t$, not as a square root.
	Note that  the second moment of the distribution \eqref{exp2f} diverges, and formally  the diffusion coefficient, corresponding to this process,  is infinite. Indeed, however, it only means that the notion of the diffusion coefficient is useless in this problem and proliferation of the charges occurs in a much faster process. The domains with different modes of relaxation on the $r-t$ plane are shown in Fig.\ref{2d}

 \subsection{Inhomogeneous relaxation in one-dimensional arrays}

 Here we have
 \begin{align}
\overline{G}_{ij}(t)\approx\int\frac{adk}{2\pi}\exp\left\{-\frac{t}{\tau}\frac{\Lambda(ka)^2|\ln(ka)|}{1+\Lambda(ka)^2|\ln(ka)|}-ikr\right\},
\label{2dd0ww}
	\end{align}
so that at the early stage of relaxation, for $t\lesssim \tau$
 \begin{align}
\overline{G}_{ij}(t)\approx tR_{\rm eff}\left[\frac{a^2d^2 U(r)}{dr^2}\right]_{r=r_{ij}}\approx\nonumber\\\approx\frac{t}{\tau}\frac{a^3}{\varepsilon_{\rm out}}\left\{\frac{2}{(r^2+r_c^2)^{3/2}}+\frac{\ln\Lambda a^3}{2r_c^3}e^{-r/r_c}\right\}\approx\nonumber\\\approx \frac{t}{\tau}\frac{1}{\varepsilon_{\rm out}}\left\{\begin{aligned}\frac{\ln\Lambda a^3}{2r_c^3}(1-r/r_c),\qquad & (a\ll r\ll r_c),\\
2a^3/r^3,\qquad & (r\gg r_c\ln\ln\Lambda),
\end{aligned}\right.
\label{gr7aea1wer}
	\end{align}
where $r_c=r_c^{(1D)}\equiv a\left(\Lambda\ln\Lambda/2\right)^{1/2}$. At the late stage, for $t\gg \tau$, the charge proliferates due to a sort of quasidiffusion:
 \begin{equation}
\label{report32_11}
\overline{G}_{ij}(t)=a(4\pi D_{\rm eff}t)^{-1/2}\exp\left\{-r^2/4D_{\rm eff}t\right\},
\end{equation}
where the ``diffusion coefficient''
 \begin{equation}
\label{report32_11a}
 D_{\rm eff}=D_{\rm eff}(r)\approx\frac{r_c^2}{\tau}\left\{1+\frac{\ln (t/\tau)}{\ln\Lambda}\right\}.
\end{equation}
depends (though, only logarithmically) on time $t$. The derivation of this result is given in Appendix \ref{Quasidiffusion in one dimension}.

\section{Conductivity in an external electric field}

When discussing the charge relaxation and deriving the expression \eqref{mr1a} and \eqref{aver-cur} for average charges and currents, we had in mind only a ``partial averaging''. Namely, we had fixed some particular spatial distribution of charges $Q_i(0)$ at time $t=0$, and then averaged over all possible evolutions of this distribution up to time $t$. Thus ``partially averaged'' distributions of charges and currents keep the memory of the initial distribution at $t=0$, though this memory is fading with time. In the absence of external field the partially averaged currents and charges relax to zero for full averaging, i.e., for $t\to\infty$. It is not the case, however, in the external field. It is only the fully averaged charges, that vanish in the presence of external field, while the fully averaged currents remain finite. Indeed, the fully averaged potentials do not vanish:
	\begin{align}
\overline{V}_i(t)=V^{(E)}_i=-({\bf E}\cdot{\bf r}_i),
\label{aver-cur1}
	\end{align}
so that the fully averaged currents are
\begin{align}
\overline{I}_{\langle nm\rangle}(t)=-R_{\rm eff}^{-1}({\bf E}\cdot\vec{\delta}_{nm}),
\label{aver-cur2}
	\end{align}
	where $\vec{\delta}_{nm}$ is a vector, connecting neighbouring sites $n$ and $m$. As a result, the conductivity of the system is
\begin{align}
\sigma=a^{2-D}R_{\rm eff}^{-1}=a^{2-D}wC_{\rm eff}.
\label{sigma}
\end{align}
We see that $\sigma$ is proportional to $C_{\rm gap}$. So, to have large conductivity, one should choose an array with narrow insulating gaps $d$ between grains and large ``contact surfaces''.

\section{Fluctuations \label{Fluctuations}}

In contrast to the currents, which do not vanish upon full averaging in the presence of an external field, the average charges are zeros even in the field. It is clear, however, that the collisions  produce fluctuations of charges; a study of these fluctuations will be our task in this section. In particular, we will find the correlators of charges and fields
\begin{align}
K(i-j,t)\equiv\overline{Q_i(t)Q_j(0)},
\label{st1a}\\
\tilde{K}_{\mu\mu'}(n-n',t)\equiv\overline{E_{\mu}(n,t)E_{\mu'}(n',0)},
	\end{align}
and demonstrate, that, in the leading approximation,
\begin{enumerate}
\item The time dependence of both $K(i-j,t)$ and $\tilde{K}_{\mu\mu'}(n-n',t)$ is trivial: it describes the homogeneous relaxation
\begin{align}
K(i-j,t)=e^{-t/\tau}K_{i-j}(0),\\ \tilde{K}_{\mu\mu'}(n-n',t)=e^{-t/\tau}\tilde{K}_{\mu\mu'}(n-n',0).
\label{st1b}
	\end{align}
	\item Characteristic amplitude of charge fluctuations 
\begin{align}
Q_0=(K_{0}(0))^{1/2}\sim C_{\rm eff}aE
	\end{align}
\item Characteristic amplitude of field fluctuations 
\begin{align}
E_\mu=(\tilde{K}_{\mu\mu}(0))^{1/2}\sim (AE^2+BE_\mu^2)^{1/2}
	\end{align}
is of the order of external field $E$. These fluctuations are anisotropic: they are stronger in the direction of the external field.
\item Correlations of the charges vanish for distances ${\bf r}_{ij}$ larger than one lattice spacing.
\item For the array dimension $D>1$ the correlations of field fluctuations decay with distance as $r^{-D}$. The sign and the magnitude of correlations strongly depend on orientation of the bonds and of the electric field.
 For $D=1$ field fluctuations at different bonds are not correlated at all.
\end{enumerate}

\subsection{General expressions for the correlator of charge fluctuations\label{General expressions for the correlator of charge fluctuations}}

To evaluate the correlator \eqref{st1a} in an external field, we write down the general solution of the inhomogeneous evolution equations \eqref{capp5}:
\begin{align}
Q_i(t)=\sum_{p=1}^{\infty}\sum_j\left[\prod_{q=1}^{p-1}\hat{g}^{(q)}\right]_{ij}X_j^{(p)},
\label{tt_2a}
\end{align}
where indices $p,q$ numerate the collisions within the interval $(-\infty,t)$ and the numeration starts from the latest collision within this interval. 
Let us first find the same-time correlator
\begin{align}
K_{ij}\equiv\overline{Q_i(t)Q_j(t)}=\nonumber\\=\sum_{\scriptsize\begin{array}{c}  p_1=1 \\ p_2=1 \end{array} }^{\infty}\overline{\left[\prod_{q_1=1}^{p_1-1}\hat{g}^{(q_1)}\right]_{ii'}X_{i'}^{(p_1)}\left[\prod_{q_2=1}^{p_2-1}\hat{g}^{(q_2)}\right]_{jj'}X_{j'}^{(p_2)}}
\label{st1}
	\end{align}
We note, that the cross terms in the sum (those with $p_1\neq p_2$) vanish after averaging. Indeed, consider, for instance, the case $p_1<p_2$. Then, since the indices $n_{p_2},m_{p_2}$ appear only in the factor $\vec{X}^{(p_2)}$, this factor turns out to be statistically independent from all other factors in this term, and should be averaged independently. On the other hand, $\overline{X}_{i}=0$, and thus the terms with $p_1\neq p_2$ fall out. Hence
\begin{align}
K_{ij}(0)=\sum_{p=1}^{\infty}\overline{\sum_{i'j'}\left[\prod_{q=1}^{p-1}\hat{\cal G}^{(q)}\right]^{ij}_{i'j'}X_{i'}^{(p)}X_{j'}^{(p)}}=\nonumber\\=\sum_{p=1}^{\infty}\sum_{i'j'}\left[\hat{\cal G}^{p-1}\right]^{ij}_{i'j'}\overline{X_{i'}X_{j'}}=\sum_{i'j'}\left[\frac{1}{1-\hat{\cal G}}\right]^{ij}_{i'j'}\overline{X_{i'}X_{j'}},
\label{st2}
	\end{align}
where 
\begin{align}
\overline{X_{i}X_{j}}=\frac{\Theta_{ij}}{N_b},\qquad \Theta_{ij}=C_{\rm eff}^2\sum_{\vec{\delta}}(\vec{\delta}\cdot\vec{E})^2(\delta_{ij}-\delta_{ij+\vec{\delta}}),
	\end{align}
 $N_b\to\infty$ being the total number of bonds in the lattice,
and the four-index object $\hat{\cal G}$ is defined as an average over the positions of the collision-bond $\langle nm\rangle$:
\begin{align}
{\cal G}^{ij}_{i'j'}\equiv\overline{\hat{g}_{ii'}^{\langle nm\rangle}\hat{g}_{jj'}^{\langle nm\rangle}}=\delta_{ii'}\delta_{jj'}+N_b^{-1}\hat{\Gamma}^{ij}_{i'j'},
\label{gam-def1}
	\end{align}
\begin{align}
\hat{\Gamma}=\hat{\Gamma}^{(0)}+\hat{\Gamma}^{(1)},
\label{gam-def2}
\end{align}  
\begin{align}
\left[\hat{\Gamma}^{(0)}\right]^{ij}_{i'j'}=C_{\rm eff}\left\{\delta_{ii'}\left[\hat{\Delta}\cdot\hat{U}\right]_{jj'}+\delta_{jj'}\left[\hat{\Delta}\cdot\hat{U}\right]_{ii'}\right\},
\label{gam-def3}
\end{align}
\begin{align}
\label{fl3}
\left[\hat{\Gamma}^{(1)}\right]^{ij}_{i'j'}=C_{\rm eff}^2\sum_{\vec{\delta}}\left(U_{j-j'}-U_{j-j'+\vec{\delta}}\right)\times\nonumber\\ \times \left\{\delta_{ij}\left(U_{i-i'}-U_{i-i'+\vec{\delta}}\right)-\delta_{i,j+\vec{\delta}} \left(U_{i-i'-\vec{\delta}}-U_{i-i'}\right)\right\}.
\end{align}
Finally, substituting \eqref{gam-def1} into \eqref{st2} we get
\begin{align}
K_{ij}(0)=-\sum_{i'j'}\left[\hat{\Gamma}^{-1}\right]^{ij}_{i'j'}\Theta_{i'j'}
	\end{align}

The two-times correlator can be found in a similar way (see Appendix \ref{two-time correlator}):
\begin{align}
K_{ij}(t)\equiv\overline{Q_i(t)Q_j(0)}=\sum_k\overline{G}_{ik}(t)K_{kj}(0)
\label{tt_1}
	\end{align}
where the average Green function $\overline{G}_{ik}(t)$ is given by \eqref{gf6}.

The four-index objects appear in our correlators because the bilinear expressions like \eqref{st2} contain two operators 	$\hat{g}^{\langle nm\rangle}$ corresponding to the same collision. They are not statistically independent and can not be averaged separately. If one would need to calculate some $N$-point correlator, one will have to deal with $2N$-index objects -- direct products of $N$ $\hat{g}$-matrices -- $\overline{\hat{g}^{\langle nm\rangle}\otimes\hat{g}^{\langle nm\rangle}\cdots \otimes\hat{g}^{\langle nm\rangle}}$. A similar situation occurs in quantum mechanics, where calculation of correlators also leads to multi-index objects -- the multi-particle Green functions.

For general $\hat{U}$ the four-index object $\hat{\Gamma}$, defined by (\ref{gam-def2},\ref{gam-def3},\ref{fl3}) is hard to deal with, because the translational invariance allows to exclude only one of the four indices. That is not enough for easy diagonalization of $\hat{\Gamma}$: after the Fourier transformation one is still left with an integral equation to solve, and its solution is not accessible for general $\hat{U}$. Similarly, in quantum mechanics, the general translational invariance allows to reduce the calculation of the two-particle Green-function to the problem of one particle in an external potential, which is not always possible to solve. It should be noted, however, that the analogy of our problem to the quantum mechanics is only partial, since our ``two-particle Hamiltonian'' $\hat{\Gamma}$ is not a hermitian one.

In the leading approximation in $1/\Lambda\ll 1$, however, this problem can be circumvented, as we will demonstrate in the next subsection.

\subsection{Correlators in the leading approximation \label{Correlators in the leading approximation}}

In the leading approximation expression for $\hat{\Gamma}^{(0)}$ can be simplified:
\begin{align}
\label{fl2}
\left[\hat{\Gamma}^{(0)}\right]^{ij}_{i'j'}=-z\delta_{ii'}\delta_{jj'}
\end{align}
It is this simplification, that helps to overcome the difficulty, mentioned in the previous subsection.

Within the total space of pairs of grains ${\cal S}\equiv \left\{(ij)\right\}$ one can single out subspace ${\cal S}_1$, containing pairs of nearest neighbours together with the pairs of two identical grains:   ${\cal S}_1\equiv \left\{(ij):i-j=\vec{\delta}\textrm{ or } \vec{0}\right\}$. This subspace is invariant with respect to the operator $\hat{\Gamma}^{(1)}$ because of the specific structure of the latter. Within the leading approximation  ${\cal S}_1$ turns out to be invariant also  with respect to the unity-like $\hat{\Gamma}^{(0)}$. Since the "initial state" $\Theta$ already belongs to the subspace ${\cal S}_1$, we conclude that the entire calculation of the correlator proceeds completely within ${\cal S}_1$, 
and, in particular, the result of this calculation -- $K_{ij}$ -- also belongs to ${\cal S}_1$. It means that the correlator does not vanish only if $i-j=\lambda$, where $\lambda=j-i$ takes $z+1$ values: either $\lambda=0$, or $\lambda=\vec{\delta}$.

Within the subspace ${\cal S}_1$ the four-index ``matrix'' $\hat{\Gamma}^{(1)}$ and two-index ``vector'' $\Theta$ are reduced to 
\begin{align}
\left[\Gamma^{(1)}\right]^{ij}_{i'j'} \rightarrow \hat{\gamma}_{\lambda\lambda'}(i-i'),\\
\left[\Theta\right]_{ij}\rightarrow \Theta_{\lambda},
\end{align} 
where
\begin{align}
\gamma_{\delta\lambda'}({\bf r})=-C_{\rm eff}^2(U_{{\bf r}+(\delta-\lambda')}-U_{{\bf r}-\lambda'})(U_{{\bf r}+\delta}-U_{\bf r}),
\label{ga1} \\
\gamma_{0\lambda'}({\bf r})=C_{\rm eff}^2\sum_{\delta} (U_{{\bf r}}-U_{{\bf r}+\delta})(U_{{\bf r}-\lambda'}-U_{{\bf r}-\lambda'+\delta}), \label{ga1l}\\
\Theta_{\lambda}=-C_{\rm eff}^2(\vec{\delta}\cdot\vec{E})^2,\qquad \Theta_{0}=2C_{\rm eff}^2E^2a^2,
\label{ga2}
\end{align}
and we get
\begin{align}
\label{Ee1}
K_{ij}\rightarrow K_{\lambda}=\sum_{\lambda'}T_{\lambda\lambda'}\Theta_{\lambda'}
\end{align}
\begin{align}
T_{\lambda\lambda'}=\left[(z-\hat{\gamma})^{-1}\right]_{\lambda\lambda'},\qquad \gamma_{\lambda\lambda'}=\sum_{\bf r}\gamma_{\lambda\lambda'}({\bf r}).
\end{align}
Evaluation of the matrices $\hat{\gamma}$ and $\hat{T}$ is presented in the Appendix \ref{diagonalization}, here we give only the final results of calculations:
\begin{align}
K_0=\frac{2z}{z^2-2}(C_{\rm eff}aE)^2,\label{finn1}\\ 
K_{\vec{\delta}}=-\left(C_{\rm eff}Ea\right)^2\left(A+Be_\mu^2\right),\label{finn2}
\end{align}
\begin{align}
A=\frac{2f_1z^2}{(z^2-2)(z^2(1+f_1)-2)},\quad B=\frac{z}{z^2(1+f_1)-2},\label{mixx1}
\end{align}
where $\mu$ denotes the Cartesian axis, parallel to $\vec{\delta}$, ${\bf e}$ is the unit vector, parallel to ${\bf E}$, and the constant $f_1$ is defined by \eqref{diagam9}, \eqref{find1}.
In one dimension we obtain:
\begin{align}
K_0=2(aC_{\rm eff}E)^2,\quad
K_{\vec{\delta}}=-(aC_{\rm eff}E)^2,
\label{cor1d0}
\end{align}
In two dimensions:
\begin{align}
K_0=\frac{4}{7}(C_{\rm eff}aE)^2,\label{cor2d0x}\\
K_{\vec{\delta}}=-\frac{2}{2+7\pi}\left(\frac{1}{7}+\pi e_\mu^2\right)(C_{\rm eff}aE)^2,\label{cor2d0}
\end{align}
In three dimensions:
\begin{align}
K_0=\frac{6}{17}(C_{\rm eff}aE)^2\label{cor3d0x}\\
K_{\vec{\delta}}\approx -(C_{\rm eff}aE)^2 (0.0026+0.1689e_\mu^2).\label{cor3d0}
\end{align}

We conclude that in the leading approximation the fluctuations of the charges are correlated only at neighbouring grains. This correlation is negative, since the collisions, being the source of fluctuations, lead to opposite charging of  the neighbouring grains that participate in the collision. We also see that the correlation for the pairs of grains, aligned along the direction of external field,  is much (in three dimensions -- almost by two orders of magnitude!) stronger, than for pairs, aligned in the perpendicular direction.

At the end of this subsection we note that in the leading approximation $G_{ii'}(t)=e^{-t/\tau}\delta_{ii'}$. So, in this approximation we obtain a  homogeneous time-decay of the two-times correlator.
\begin{align}
K_{ij}(t)=e^{-|t|/\tau}K_{ij}(0)
\end{align}

\subsection{Correlator of local electric fields\label{Correlators of currents}}

\begin{figure}[h]
\includegraphics[width=0.7\columnwidth]{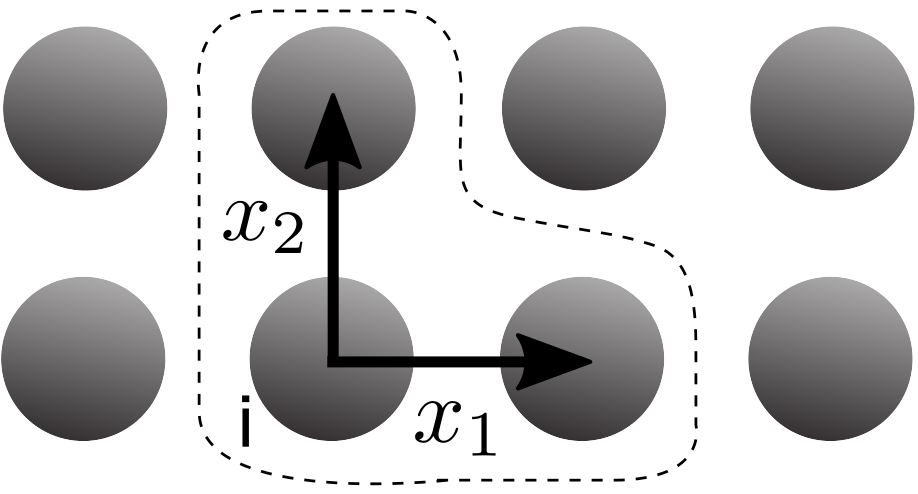}
\caption{The unit cell for the case of 2D square array contains one site and two bonds.} \label{grid}
\end{figure}

The local potential drops $V_{\langle nm\rangle}\equiv V_n-V_m$ (related to the electric fields $E_{\langle nm\rangle}=-(V_n-V_m)/a$) are associated with particular bonds $\langle nm\rangle$. There are $D=z/2$ different bonds in the unit cell (see Fig. \ref{grid}), therefore we will mark the bonds by two indices:  by the coordinate ${\bf r}_i$ of the reference site in the cell, and by spatial directions $\mu=x_1,x_2,\ldots x_D$ of the bonds. 

Then the same-time correlator of fields
\begin{align}
\tilde{K}_{\mu\mu'}(n-n')\equiv 
\overline{\delta E_{\langle n,n+\vec{\delta}_{\mu}\rangle}\delta E_{\langle {n',n'+\vec{\delta}_{\mu'}\rangle}}}=\nonumber\\=\sum_{i\lambda}(U_{n-i}-U_{n+\vec{\delta}_{\mu}-i})(U_{n'-i-\lambda}-U_{n'+\vec{\delta}_{\mu'}-i-\lambda})\frac{K_\lambda}{a^2}=\nonumber\\=\left(\frac{z}{4aC_{\rm eff}}\right)^2\int_{\rm BZ}\frac{a^Dd^D{\bf k}}{(2\pi)^D}e^{i({\bf k}\cdot {\bf r}(n\mu| n'\mu'))}K({\bf k})\nonumber\\\times \frac{\sin(ak_{\mu}/2)\sin(ak_{\mu'}/2)\sin^2(ak_{\mu''}/2)}{\left[\sum_\nu\sin^2(ak_{\nu}/2)\right]^2},
\end{align}
where $\vec{\delta}_{\mu}$ is a vector of length $a$ in the positive direction of the axis $\mu$ and
\begin{align}
{\bf r}(n\mu| n'\mu'))\equiv ({\bf r}_n-{\bf r}_{n'})+\frac12(\vec{\delta}_{\mu}-\vec{\delta}_{\mu'})
\end{align}
is nothing else, but the vector, connecting centres of the bonds $(n\mu)$ and  $(n'\mu')$. 
\begin{align}
K({\bf k})=\sum_{\lambda}K_{\lambda}e^{i({\bf k}\cdot \lambda)}=-\sum_{\vec{\delta}}K_{\vec{\delta}}(1-e^{i({\bf k}\cdot \vec{\delta})})=\nonumber\\=4\left(C_{\rm eff}Ea\right)^2\sum_{\mu}\left(A+Be_\mu^2\right)\sin^2(ak_{\mu}/2)
\end{align}
\begin{align}
\tilde{K}_{\mu\mu'}(n-n')=\left(\frac{zE}{2}\right)^2\sum_{\mu''}\left(A+Be_\mu''^2\right)O_{\mu\mu'\mu''}({\bf r}(n\mu| n'\mu'))\end{align}
\begin{align}
O_{\mu\mu'\mu''}({\bf r})=\int_{\rm BZ}\frac{a^Dd^D{\bf k}}{(2\pi)^D}e^{i({\bf k}\cdot {\bf r})}\nonumber\\\times\frac{\sin(ak_{\mu}/2)\sin(ak_{\mu'}/2)\sin^2(ak_{\mu''}/2)}{\left[\sum_\nu\sin^2(ak_{\nu}/2)\right]^2}
\end{align}
For the same-bond correlator of fields we have
\begin{align}
\overline{\delta E^2_{\mu}}=\tilde{K}_{\mu\mu}(0)=-\left(\frac{zE}{2}\right)^2\sum_{\mu''}\left(A+Be_\mu''^2\right)O_{\mu\mu\mu''}(0)=\nonumber\\=E^2\left(\tilde{A}+\tilde{B}e_\mu^2\right),
\label{werr1}
\end{align}
where
\begin{align}
\tilde{A}=\left(\frac{z}{2}\right)^2\frac{2f_1z^3}{(z^2(1+f_1)-2)(z^2-2)}
,\nonumber\\
\tilde{B}=\left(\frac{z}{2}\right)^2\frac{2-f_1z^2}{z^2(1+f_1)-2},\label{mixx2}
\end{align}
 The calculation of the constants $\tilde{A},\tilde{B}$ is given in Appendix \ref{Asymptote of the correlator of fields}. In particular, for the 2D-array
 \begin{align}
\tilde{A}=\frac{32}{7}\frac{1}{2+7\pi},\qquad
\tilde{B}=4\frac{\pi-2}{2+7\pi}.\nonumber
\end{align}

 Let us now turn to the correlations of fields at different bonds.
In one-dimensional case there is only one option for all three indices $\mu,\mu',\mu''=x$, so that
\begin{align}
O_{xxx}(x)\equiv O(x_{nn'})=\int_{\pi/a}^{\pi/a}\frac{adk}{2\pi}e^{ikx_{nn'}}=\delta_{nn'},
\end{align}
\begin{align}
\tilde{K}(n-n')=E^2(A+B)\delta_{nn'}
\end{align}

Thus, in 1D the fluctuations of fields are not correlated in space. The physical reason for that is clear: There is only one bond in the unit cell of the one-dimensional chain, and the potential drops $V_{\langle nm\rangle}$ can be chosen as {\it independent variables} instead of charges $Q_i$. Since, within the leading approximation in $\Lambda$, the Coulomb energy of the system is diagonal in these variables, their fluctuations are statistically independent.

In dimensions $D>1$ the correlations of fields at different bonds do not vanish.
In particular, for the case $n-n'\gg 1$, we obtain
\begin{align}
\tilde{K}_{\mu\mu'}(n-n')\approx\frac{(zE/2)^2}{(r_{nn'}/a)^D}\nonumber\\\times\left\{\delta_{\mu\mu'}\left[a+be_{\mu}e_{\mu'}+c\sum_{\nu}n_{\nu}^2e_{\nu}^2\right]-\right.\nonumber\\-\left.\frac{z}{2}n_{\mu}n_{\mu'}\left[a+b(e_{\mu}^2+e_{\mu'}^2)+c\left(1+\frac{4}{z}\right)\sum_{\nu}n_{\nu}^2e_{\nu}^2\right]\right\}
\label{srett1}
\end{align}
Derivation of the results \eqref{srett1} and general expressions for the constants $a,b,c$ are presented in the Appendix \ref{Asymptote of the correlator of fields}. Here we give only the values of constants for the 2D-array:
\begin{align}
a=\frac{1}{14\pi}\approx 0.0227,\qquad 
 b=-c=\frac{1}{2+7\pi}\approx 0.0417.\nonumber
\end{align}

\begin{figure}
\includegraphics[width=1\columnwidth]{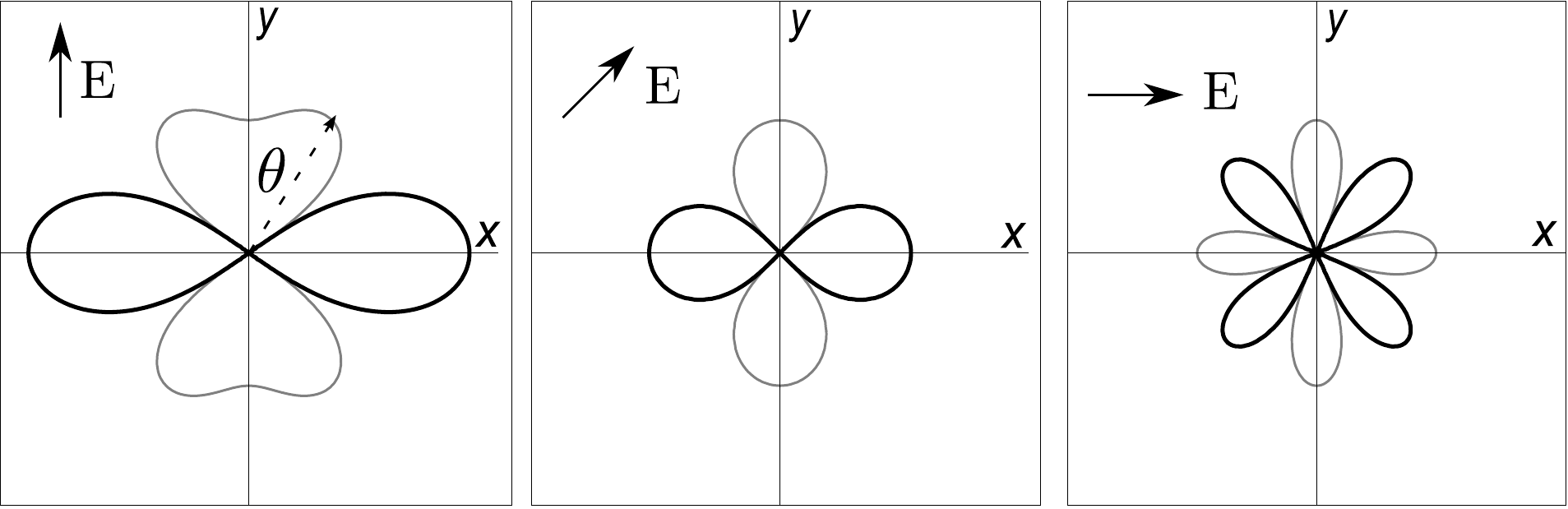}
\caption{The angular dependence of the correlator of electric  fields for a two-dimensional square lattice of grains. Both bonds are assumed to be parallel to each other and to the axis $y$. The length of the radius-vector is proportional to $|\tilde{K}_{yy}({\bf r})|$, while the angle $\theta$ represents the angle between ${\bf r}$ and the crystallographic $y$-axis. Direction of the external field is indicated by the arrow. Correlation is positive (negative) at the lobes, depicted with thick (thin) lines} \label{corr-diagr}
\end{figure}

The angular dependence of the correlator of electric  fields on parallel bonds is analysed in Fig.\ref{corr-diagr}.

\section{Limitations of the model \label{Limitations of the model}}

The most principal limitation of the approach, presented in this paper, is the ``classical approximation'', described in the Section \ref{Classic approximation} -- the assumption of the continuous charge. It is this assumption, that allows for finding the charge distribution after each collision from the condition of minimal energy \eqref{coulomb2aa} without applying the constraint of the charge discreteness.  Obviously, this approximation can be justified if, due to some reason, the fluctuations of charges are large compared to $e$. Using the results of Section \ref{Correlators in the leading approximation} (see \eqref{cor1d0},\eqref{cor2d0x},\eqref{cor3d0x}), we see, that this condition is fulfilled, in particular, if the electric field, applied to the system, is large enough: 
\begin{align}
\label{lim1}
E\gg e(aC_{\rm eff})^{-1}.
\end{align} 

 Being taken into account, charge discreteness leads to  the Coulomb blockade effect. The randomness of the offset charges $q_i$, in composition with the Coulomb blockade leads to strong suppression of both electronic transport and relaxation in a granular system, so that ``variable range cotunnelling''\cite{hopping} becomes the leading  mechanism of conductivity at low temperatures. In nonstationary system, however, the offset charges are subject to temporal fluctuations, and these fluctuations may locally suppress the Coulomb blockade and thus facilitate transport. This interesting effect we will discuss in a separate publication.

Another important factor, which we did not take into account in this paper, is temperature. For finite $T$ one has to take into account thermal fluctuations governed by Gibbs distribution, and the condition that the energy should be minimized after each collision is, strictly speaking, not valid. 
Thermal fluctuations in the system should be treated in parallel with the external nonthermal fluctuations, considered in this paper. The most obvious effect of thermal fluctuations would be $T$-dependent renormalization of the fluctuations rate $w$. However, some other less trivial effects are also expected.

\section{Conclusion \label{Conclusion}}

We have considered transport processes in an array of conducting grains, connected by nonstationary resistances. These processes  consist of a sequences of local breakdowns, occurring due to ``collisions'' of grains with each other. Three-dimensional, as well as low-dimensional arrays were studied. We were able to find conductivity, relaxation rate and correlators of fluctuations (of local charges and local fields) in the ``classical approximation'', which neglects the discreteness  of  charges. This approximation is valid, if the charge fluctuations are large due to strong applied fields and/or  to large capacitances of the intergrain contacts. The effects of the charge discreteness, essential for the low-field conductivity will be discussed in a separate publication.

We have studied in detail the case, when the electrostatic properties of the system are dominated by the intergrain capacitances $C_{\rm gap}$, which is true for a system with narrow dielectric gaps between grains. Both conductivity of the system and the amplitude of fluctuations appear to be proportional to $C_{\rm gap}$. The correlations of the charge fluctuations are of a short range type, while the spatial correlations of electric fields decay as a power law, and are strongly anisotropic.

Possible candidates for the realization of the model were discussed: the colloidal solutions, and the polymer-linked systems of metal grains, including the arrays of nanomechanical shuttles.

We are indebted to M.V.Feigel'man, L.B.Ioffe, E.I.Kats, V.E.Kravtsov, and D.S.Lyubshin for helpful comments and advises. Special thanks are due to G.A.Tsirlina, who has guided us in the field of the colloid science and chemistry of nanoparticles. This work was supported by 5top100 grant of Russian Ministry of Education and Science.

\appendix


\section{Evaluation of $U_{ij}$ in one-dimensional case\label{Evaluation of in one-dimensional case}}

Using exact relation between $U({\bf r})$ and $C({\bf k})$ (the upper line of \eqref{green1pw}) together with the definitions \eqref{correction-k1}, we can write
 \begin{align}
 U(r)=\frac{1}{C_{\rm gap}}\int_{-\pi/a}^{\pi/a}\frac{adk}{2\pi}\frac{e^{-ikr}}{(ka)^2-\frac{1}{\Lambda\ln(|k|a)}}=\nonumber\\=\frac{1}{\varepsilon_{\rm out}r_c^{(1D)}}F^{(1D)}(r/r_c^{(1D)},\ln\Lambda),
\label{correction-k1w}
	\end{align}
where
\begin{align}
 r_c^{(1D)}=a\left(\Lambda\ln\Lambda/2\right)^{1/2},
\\
F(z,y)=\frac{y}{\pi}\int_0^{\infty}\frac{\cos z\xi d\xi}{\xi^2+\left(1-\frac{2 \ln\xi}{y}\right)^{-1}}, \label{green1q18}
	\end{align}
and we are interested in  behaviour of this function for arbitrary $z$ and large $y\gg 1$.

For $z\lesssim 1$ one can neglect the logarithmic correction in the denominator, so that
\begin{align}
F(z,y)\approx\frac{y}{\pi}\int_0^{\infty}\frac{\cos z\xi d\xi}{\xi^2+1}=\frac{y}{2}e^{-z},\qquad (z\lesssim 1).
\label{green1q18k}
	\end{align}

To evaluate the integral \eqref{green1q18} for $z\gg 1$  it is convenient to decompose the integrand in three parts:
\begin{align}
\frac{1}{\xi^2+\left(1-\frac{2 \ln\xi}{y}\right)^{-1}}\equiv \frac{1}{\xi^2+1}-\frac{2\ln\xi}{y}-\nonumber\\-\frac{2\xi^2\ln\xi}{y}\frac{1+(\xi^2+1)\left(1-\frac{2 \ln\xi}{y}\right)}{(\xi^2+1)\left[\xi^2\left(1-\frac{2 \ln\xi}{y}\right)+1\right]}
\label{green1q18d}
	\end{align}
	\begin{enumerate}
\item The first term in \eqref{green1q18d} after the integration gives exactly the result \eqref{green1q18k}, coming from $\xi\sim 1$.
\item The second term, after integration by parts and regularization, gives
\begin{align}
-\frac{2}{\pi }\int_0^{\infty}\cos z\xi d\xi\ln \xi=\frac{2}{\pi z}\int_0^{\infty}\frac{\sin z\xi}{\xi}d\xi=\frac{1}{z},
\label{asy3}
	\end{align}
This contribution comes mainly from  $\xi\sim z^{-1}\ll 1$.
\item 	The third term gives small contribution $\sim 1/z^{3}$, that can be neglected
\end{enumerate}
 Thus, for $z\gg 1$ there are two independent contributions to \eqref{green1q18}, coming from two different domains of $\xi$ that do not overlap for $z\gg 1$:
\begin{align}
F(z,y)=\frac{1}{z}+\frac{y}{2}e^{-z}.
\label{asy9}
	\end{align}
The first term in \eqref{asy9} dominates for $z\gg\ln y$, while the second one dominates at $1\ll z\ll \ln y$. As we have seen, the second term in \eqref{asy9} gives a correct result also for $z\lesssim 1$. If the second term  dominated over the first one for all $z\lesssim 1$, we would use the result \eqref{asy9}  for all $z$ -- small, large and intermediate. Unfortunately, however, it is not possible: for very small $z<1/y$ the first term begins to dominate again and  \eqref{asy9} becomes incorrect for these small $z$. To get rid of this problem it is  enough just to regularise the first term so that it ceases to  grow at $z\lesssim 1$. For example, one can choose
	\begin{align}
F(z,y)=\frac{1}{\sqrt{z^2+1}}+\frac{y}{2}e^{-z}
\label{asy9g}
	\end{align}
Other regularisations are also possible: different ones may be not equivalent from the practical point of view (quality of description of numerical data at finite $y$ may be different), but all of them are legitimate since they all should be asymptotically correct at $y\gg1$.

\section{Quasidiffusion in one dimension \label{Quasidiffusion in one dimension}}

At late stage of the evolution the charge is already far from the origin, and the characteristic value $k_0$ of momentum $k$ in the integral \eqref{2dd0ww} are so small that $\Lambda(ka)^2|\ln(ka)|\ll 1$ and we can write
 \begin{align}
\overline{G}_{ij}(t)\approx\int\frac{dk}{2\pi}\exp\left\{-\frac{t}{\tau}\Lambda(ka)^2|\ln(ka)|-ikr\right\}.
\label{2dd0wwe}
	\end{align}
The momentum $k$ appearing under the logarithm in \eqref{2dd0wwe} can be substituted by its characteristic value $k_0$ and we get
\begin{align}
\overline{G}_{ij}(t)\approx\int\frac{dk}{2\pi}\exp\left\{-tD_{\rm eff}k^2-ikr\right\},\label{2dd0wwe1q}\\
D_{\rm eff}=\frac{\Lambda a^2}{\tau}|\ln(k_0a)|=\frac{r_c^2}{\tau}\frac{2|\ln(k_0a)|}{\ln\Lambda}.
\label{2dd0wwe1}
	\end{align}
Now we have to estimate $k_0=k_0(t,r)$ in the Gaussian integral \eqref{2dd0wwe1q}: 
\begin{align}
k_0(t,r)\sim\max\left\{(tD_{\rm eff})^{-1/2},\; r/(tD_{\rm eff})\right\}.
\label{2dd0wwe2}
	\end{align}
Substituting this result into \eqref{2dd0wwe1}, and neglecting the $\log\log$-corrections, we get
\begin{align}
D_{\rm eff}=\frac{2r_c^2}{\tau\ln\Lambda}\left|\ln\max\left\{[(t/\tau)\Lambda]^{-1/2},\; (\tau/t)(ar/r_c^2)\right\}\right|.
\label{2dd0wwe3}
	\end{align}
We are mostly interested in the spatial domain $r\lesssim \overline{r}(t)=(tD_{\rm eff})^{1/2}$, where almost all the charge is confined. In this domain the maximum in \eqref{2dd0wwe3} is dominated by the first term, and we finally arrive at the result \eqref{report32_11a}. One should have in mind, however, that for large $r\gg \overline{r}(t)$ the effective diffusion coefficient logarithmically depends not only on $t$, but also on $r$.

\section{Calculation of the two-times correlator \label{two-time correlator}}

For the two-time correlator
we need
\begin{align}
Q_j(0)=\sum_{p_2=1}^{\infty}\sum_{j'}\left[\prod_{q_2=1}^{p_2-1}\hat{g}^{(q_2)}\right]_{jj'}X_{j'}^{(p_2)},\\
Q_i(t)=\sum_{p_1=1}^{\infty}\sum_{ki'}\left[\prod_{s=1}^{M}\hat{g}^{(s)}\right]_{ik}\left[\prod_{q_1=1}^{p_1-1}\hat{g}^{(q_1)}\right]_{ki'}X_{i'}^{(p_1)}+\nonumber\\+\sum_{u=1}^{M}\sum_{i'}\left[\prod_{s=1}^{u-1}\hat{g}^{(s)}\right]_{ii'}X_{i'}^{(u)}.
\label{tt_2}
	\end{align}
Here indices $q,p$ numerate breakdowns within the interval $(-\infty,0)$ and indices $s,u$ -- within the interval $(0,t)$. Both numerations start from the latest breakdown within the corresponding interval. 
It is easy to understand that 
\begin{enumerate}
\item The terms, containing $X_{i'}^{(u)}$ vanish upon averaging of $Q_i(t)Q_j(0)$ in \eqref{tt_1}.
\item The factors $\hat{g}^{(s)}$ coming from the first term in \eqref{tt_2} are statistically independent both with respect to each other and with respect to all other factors in the product.
\item The cross terms with $p_1\neq p_2$ vanish in the same way, as it happened in \eqref{st1}.
\end{enumerate}
As a result, we arrive at
\begin{align}
K_{ij}(t)=\sum_k\overline{\left[\prod_{s=1}^{M}\hat{g}^{(s)}\right]}_{ik}\times\nonumber\\\times\sum_{p=1}^{\infty}\overline{\left[\prod_{q_1=1}^{p-1}\hat{g}^{(q_1)}\right]_{ki'}\sum_{p=1}^{\infty}\left[\prod_{q_2=1}^{p-1}\hat{g}^{(q_2)}\right]_{jj'}}\overline{X_{i'}^{(p)}X_{j'}^{(p)}},
\label{tt_3}
	\end{align}
which can be rewritten in the form \eqref	{tt_1}.


\section{Evaluation of the matrices $\hat{\gamma}$ and $\hat{T}$\label{diagonalization}}

As it follows from \eqref{ga1}, \eqref{ga1l},
\begin{align}
\sum_{\lambda}\gamma_{\lambda\lambda'}=0,\label{ga1wds} 
\end{align}
This property allows, as the first step, to exclude the components with $\lambda=0$ from all the objects that we are dealing with. In particular,
\begin{align}
\label{Ee1azz}
 K_{\vec{\delta}}=\sum_{\vec{\delta}'}\tilde{T}_{\vec{\delta}\vec{\delta}'}\Theta_{\vec{\delta}'},\quad K_0=-\sum_{\vec{\delta}}K_{\vec{\delta}}, \\\tilde{T}_{\vec{\delta}\vec{\delta}'}=\left[(z-\hat{\tilde{\gamma}})^{-1}\right]_{\vec{\delta}\vec{\delta}'},\qquad
\tilde{\gamma}_{\vec{\delta}\vec{\delta}'}=\gamma_{\vec{\delta}\vec{\delta}'}-\gamma_{\vec{\delta}0}.\label{yy}
\end{align}
 It is much more convenient to work with a symmetric $z\times z$ matrix $\hat{\tilde{\gamma}}$, than with the initial asymmetric $(z+1)\times (z+1)$ matrix $\hat{\gamma}$.

 There are only two different matrix elements in the matrix $\hat{\tilde{\gamma}}$. For $\vec{\delta}\neq\pm\vec{\delta}'$
\begin{align}
\tilde{\gamma}_{\vec{\delta}\vec{\delta}'}= \int_{\rm BZ}\frac{(1-\cos ak_x)(1-\cos ak_y)}{2\left[\sum_{\mu'}(1-\cos ak_{\mu'})\right]^2}\frac{a^Dd^D{\bf k}}{(2\pi)^D}=f_1,\label{diagam9}
\end{align}
 and
\begin{align}
\tilde{\gamma}_{\vec{\delta}\vec{\delta}}=\tilde{\gamma}_{\vec{\delta},-\vec{\delta}}=\int_{\rm BZ}\frac{(1-\cos ak_x)^2}{2\left[\sum_{\mu'}(1-\cos a k_{\mu'})\right]^2}\frac{a^Dd^D{\bf k}}{(2\pi)^D}=\nonumber\\=f_2=\frac{1}{z}-\left(\frac{z}{2}-1\right)f_1,\label{diagam10}
\end{align}
The integral \eqref{diagam9} can be calculated analytically in two dimensions, while in three dimensions it can be done only numerically:
\begin{align}
f_1^{(\rm D=2)}=1/4\pi,\qquad f_1^{(\rm D=3)}=0.0421.
\label{find1}
\end{align}

There are three groups of eigenvectors $\psi^{(\alpha)}_{\vec{\delta}}$ and corresponding eigenvalues $\tilde{\gamma}^{(\alpha)}$ of the matrix $\hat{\tilde{\gamma}}$:
\begin{enumerate}
\item $z/2$ degenerate modes, antisymmetric with respect to reflections:
\begin{align}
\psi^{(\mu-)}_{\vec{\delta}}=\chi^{(\mu-)}_{\vec{\delta}},\quad \tilde{\gamma}^{(\mu-)}\equiv\tilde{\gamma}^{(-)}=0,
\end{align}
where
\begin{align}
\chi^{(\mu\pm)}_{\vec{\delta}}=\frac{1}{\sqrt2}\left(\delta_{\vec{\delta},+\mu}\pm\delta_{\vec{\delta},-\mu}\right),
\label{diagam9er9}
\end{align}
and the indices  $\mu=x_1,x_2,\ldots x_D$ denote $D=z/2$ spatial Cartesian axes. 
\item One fully-symmetric mode
\begin{align}
\psi^{(\rm s)}_{\vec{\delta}}=\frac{1}{\sqrt{z}},\quad \tilde{\gamma}^{(\rm s)}=2f_2+(z-2)f_1=\frac{2}{z},
\end{align}
\item $z/2-1$ degenerate modes, symmetric with respect to reflections, but not fully symmetric with respect to permutations:
\begin{align}
\psi^{(\eta)}_{\vec{\delta}}=\sum_{\mu}e_{\mu}^{(\eta)}\chi^{(\mu+)}_{\vec{\delta}},\\ \tilde{\gamma}^{(\eta)}\equiv\tilde{\gamma}^{(\rm as)}=2(f_2-f_1)=\frac{2}{z}-zf_1,
\end{align}
where $z/2-1$ unit vectors $e_{\mu}^{(\eta)}$ form, together with the fully-symmetric vector $a_{\mu}^{(\rm s)}=\sqrt{2/z}$, the full orthonormal basis in the $z/2=D$-dimensional space.
\end{enumerate}
Consequently, there are three different eigenvalues for the matrix $\hat{T}$:
\begin{align}
\tilde{T}^{\rm (-)}=\frac{1}{z},\qquad \tilde{T}^{(\rm s)}=\frac{1}{z-\tilde{\gamma}^{(\rm s)}}=\frac{z}{z^2-2},\\ \tilde{T}^{\rm (as)}=\frac{1}{z-\tilde{\gamma}^{(\rm as)}}=\frac{z}{z^2(1+f_1)-2}.
\label{diagam10w}
\end{align}
It is convenient also to split $\Theta$ into three parts of the same type:
\begin{align}
\Theta^{\rm (-)}=0,\qquad\Theta^{\rm (s)}=\frac{1}{z}\sum_{\vec{\delta}}\Theta_{\vec{\delta}}=-\frac{2(C_{\rm eff}aE)^2}{z},
\end{align}
\begin{align}
\Theta^{\rm (as)}_{\vec{\delta}}\equiv \Theta_{\vec{\delta}}-\Theta^{\rm (s)}=-C_{\rm eff}^2\left((\vec{\delta}\cdot\vec{E})^2-\frac{2(Ea)^2}{z}\right)\nonumber,
\end{align}
Then
\begin{align}
K_0=-z\tilde{T}^{(\rm s)}\Theta^{(\rm s)},\qquad K_{\vec{\delta}}=\tilde{T}^{(\rm s)}\Theta^{(\rm s)}+\tilde{T}^{(\rm as)}\Theta^{(\rm as)},\nonumber
\end{align}
and we finally arrive at the final result \eqref{finn1}, \eqref{finn2}.

    \section{Derivation of the correlator of electric fields \label{Asymptote of the correlator of fields}}

For the kernel $O_{\mu\mu\mu''}(0)$, entering the same-bond correlator of fields, we have
\begin{align}
O_{\mu\mu\mu''}(0)=\int_{\rm BZ}\frac{a^Dd^D{\bf k}}{(2\pi)^D}\frac{\sin^2(ak_{\mu}/2)\sin^2(ak_{\mu''}/2)}{\left[\sum_\nu\sin^2(ak_{\nu}/2)\right]^2}.
\end{align}
Comparing this expression with \eqref{diagam9} and \eqref{diagam10}, we have
\begin{align}
O_{\mu\mu\mu''}(0)=2f_1(1-\delta_{\mu\mu''})+2\left[\frac{1}{z}-\left(\frac{z}{2}-1\right)f_1\right]\delta_{\mu\mu''}
\end{align}
and, substituting this expression to \eqref{werr1}, we arrive at
\begin{align}
\overline{\delta E^2_{\mu}}=
E^2\left(\tilde{A}+\tilde{B}e_\mu^2\right),\nonumber\\\tilde{A}=\left(\frac{z}{2}\right)^2\left[f_1(zA+2B)-\frac{f_1z^2-2}{z}A\right]
,\nonumber\\
\tilde{B}=\frac{2-f_1z^2}{z}B=\left(\frac{z}{2}\right)^2\frac{2-f_1z^2}{z^2(1+f_1)-2},
\end{align}
 from where, using expressions \eqref{mixx1} for $A$ and $B$, we  get the final result \eqref{mixx2}.

To find the kernel for the correlations of fields at bonds at large distance $r\gg a$ from each other, one should expand the integrand in small ${\bf k}$:
\begin{align}
O_{\mu\mu'\mu''}({\bf r})\approx\int_{\rm BZ}\frac{a^Dd^D{\bf k}}{(2\pi)^D}e^{i({\bf k}\cdot {\bf r})}\frac{k_{\mu}k_{\mu'}k^2_{\mu''}}{k^4}.
\label{sdeq1}
\end{align}
To evaluate the integral in \eqref{sdeq1} we consider
\begin{align}
I_{\mu_1\mu_2\mu_3\mu_4}({\bf u})=\left\langle(m_{\mu_1}m_{\mu_2}m_{\mu_3}m_{\mu_4}e^{i({\bf m}\cdot {\bf u})}\right\rangle_{\bf m}=\nonumber\\=\frac{\partial^4\varphi({\bf u})}{\partial u_{\mu_1}\partial u_{\mu_2}\partial u_{\mu_3}\partial u_{\mu_4}},
\label{cor-fil1}
\end{align}
where the averaging is performed over the unit vector ${\bf m}$. After simple calculations we get
\begin{align}
I_{\mu_1\mu_2\mu_3\mu_4}({\bf u})=\nonumber\\=\left\{\delta_{\mu_1\mu_2}\delta_{\mu_3\mu_4}+\mbox{permutations}\right\}\left(\frac{d}{udu}\right)^2\varphi+\nonumber\\+\left\{\delta_{\mu_1\mu_2}u_{\mu_3}u_{\mu_4}+\mbox{permutations}\right\}\left(\frac{d}{udu}\right)^3\varphi+\nonumber\\+u_{\mu_1}u_{\mu_2}u_{\mu_3}u_{\mu_4}\left(\frac{d}{udu}\right)^4\varphi,\nonumber
\end{align}
where
\begin{align}
\varphi({\bf u})=\left\langle e^{i({\bf m}\cdot {\bf u})}\right\rangle_{\bf m}=J_{D/2}(u)=\left\{\begin{aligned}\sin u/u,&\qquad (D=3),\\J_0(u),
&\qquad(D=2).
\end{aligned}\right.
\nonumber
\end{align}
The kernel $O_{\mu\mu'\mu''}({\bf r})$ is related to the integral $I_{\mu_1\mu_2\mu_3\mu_4}({\bf u})$:
\begin{align}
O_{\mu\mu'\mu''}({\bf r})=\frac{1}{r^D}\int \frac{u^{D-1}du}{2\pi^{D-1}}I_{\mu\mu'\mu''\mu''}({\bf n}u)=\nonumber\\=\frac{\delta_{\mu\mu'}}{r^D}\left\{I_1(1+\delta_{\mu\mu''}+\delta_{\mu'\mu''})+I_2n_{\mu''}^2\right\}+\nonumber\\+\frac{n_{\mu}n_{\mu'}}{r^D}\left\{I_2(1+2\delta_{\mu\mu''}+2\delta_{\mu'\mu''})+I_3n^2_{\mu''}\right\}\label{sdeq2}
\end{align}
where ${\bf n}\equiv {\bf r}/r$ and 
\begin{align}
I_k=\int_0^\infty \frac{u^{D-1+2(k-1)}du}{2\pi^{D-1}}\left(\frac{d}{udu}\right)^{k+1}\varphi(u).\label{cor-fil10}
\end{align}
As a result, we arrive at \eqref{srett1} with
\begin{align}
a_1=A[I_1(z/2+2)+I_2]+BI_1,\nonumber\\
b_1=2BI_1,\qquad c_1=I_2B,\nonumber\\
a_2=A[I_2(z/2+4)+I_3]+BI_2,\nonumber\\
b_2=2BI_2,\qquad c_2=I_3B,\nonumber
\end{align}
 The integrals \eqref{cor-fil10} are ultraviolet-divergent. However, their divergent parts only contribute to the short-range part of the correlator, which is zero outside the immediate vicinity of $r=0$. Indeed, consider, for example, the most divergent term $\{$m.d.t.$\}$ for the three-dimensional case. It arises from the integral $I_3$ and corresponds to the term with the highest power of $u$ in the integrand:
 \begin{align}
I_3=\int \frac{u^{6}du}{2\pi^{2}}\left(\frac{d}{udu}\right)^{4}\varphi(u)\rightarrow\int \frac{u^{2}du}{2\pi^{2}}\varphi(u)\rightarrow\delta ({\bf r})
\nonumber
\end{align}
Obviously, this term does not contribute to the long-range correlator.

 Then, after integration by parts in \eqref{cor-fil10} and discarding the divergent terms, we obtain 
\begin{align}
I_1=\left\{\begin{aligned}\int_0^\infty \frac{du}{2\pi^2}\frac{u-\sin u}{u^3}=1/8\pi,&\qquad (D=3),\\
-J_0''(0)/2\pi=1/4\pi,&\qquad (D=2),
\end{aligned}\right.
\label{cor-fil5we}
\end{align}
\begin{align}
I_2=-DI_1,\qquad
I_3=D(D+2)I_1
\label{cor-fil7}
\end{align}
and, consequently,
\begin{align}
a_1=I_1(2A+B),\quad
b_1=2I_1B,\quad c_1=-(z/2)I_1B,\nonumber\\
a_2=-\frac{zI_1}{2}(2A+B),\quad
b_2=-zI_1B,\quad c_2=\frac{z(z+4)}{4}I_1B.\nonumber
\end{align}
Substituting \eqref{mixx1} for $A$ and $B$, we  arrive at the final result:
\begin{align}
a_1=I_1(2A+B)=\frac{z(z^2+4f_1z-2)I_1}{(z^2-2)(z^2(1+f_1)-2)},\nonumber\\
b_1=\frac{2zI_1}{z^2(1+f_1)-2},\qquad c_1=-\frac{z^2I_1}{2(z^2(1+f_1)-2)},\nonumber\\
a_2=I_2(2A+B)=-(z/2)a_1,\nonumber\\
b_2=-(z/2)b_1,\qquad c_2=-[(z+4)/2]c_1,\label{srett2}
\end{align}
with the constant $I_1=(1/\pi)2^{-z/2}$.


\end{document}